\documentclass[12pt]{article}
\usepackage{epsfig}
\usepackage{psfrag}
\usepackage{latexsym}
\def\beq{\begin{equation}}
\def\eeq{\end{equation}}
\def\bea{\begin{eqnarray}}
\def\eea{\end{eqnarray}}
\def\bet{\begin{tabular}}
\def\eet{\end{tabular}}
\def\bes{\begin{subequations}\bea}
\def\ees{\eea\end{subequations}}

\newcommand{\bi}{\begin{itemize}}
\newcommand{\ei}{\end{itemize}}
\def\be{\begin{equation}}
\def\ee{\end{equation}}
\def\bc{\begin{center}}
\def\ec{\end{center}}
\def\bea{\begin{eqnarray}}
\def\eea{\end{eqnarray}}

\def\nn{\nonumber}

\catcode`@=11
\def\marginnote#1{}
\newcount\hour
\newcount\minute
\newtoks\amorpm
\hour=\time\divide\hour by60
\minute=\time{\multiply\hour by60 \global\advance\minute by-\hour}
\edef\standardtime{{\ifnum\hour<12 \global\amorpm={am}%
        \else\global\amorpm={pm}\advance\hour by-12 \fi
        \ifnum\hour=0 \hour=12 \fi
        \number\hour:\ifnum\minute<10 0\fi\number\minute\the\amorpm}}
\edef\militarytime{\number\hour:\ifnum\minute<10 0\fi\number\minute}
\def\draftlabel#1{{\@bsphack\if@filesw {\let\thepage\relax
   \xdef\@gtempa{\write\@auxout{\string
      \newlabel{#1}{{\@currentlabel}{\thepage}}}}}\@gtempa
   \if@nobreak \ifvmode\nobreak\fi\fi\fi\@esphack}
        \gdef\@eqnlabel{#1}}
\def\@eqnlabel{}
\def\@vacuum{}
\def\draftmarginnote#1{\marginpar{\raggedright\scriptsize\tt#1}}
\def\draft{\oddsidemargin 0.0truein
        \def\@oddfoot{\sl preliminary draft \hfil
        \rm\thepage\hfil\sl\today\quad\militarytime}
        \let\@evenfoot\@oddfoot \overfullrule 3pt
        \let\label=\draftlabel
        \let\marginnote=\draftmarginnote
   \def\@eqnnum{(\theequation)\rlap{\kern\marginparsep\tt\@eqnlabel}%
\global\let\@eqnlabel\@vacuum}  }
\catcode`@=12
\begin{document}
\begin{titlepage}
\vspace*{-1cm}
\phantom{hep-ph/******} 
\hfill{RM3-TH/08-14}
\hfill{CERN-PH-TH/2008-189}

\vskip 2.5cm
\begin{center}
\mathversion{bold}
{\Large\bf CP violation in neutrino oscillations and new physics }
\mathversion{normal}
\end{center}
\vskip 0.2  cm
\vskip 0.5  cm
\begin{center}
{\large Guido Altarelli}~\footnote{e-mail address: altarelli@fis.uniroma3.it}
\\
\vskip .1cm
Dipartimento di Fisica `E.~Amaldi', Universit\`a di Roma Tre
\\ 
INFN, Sezione di Roma Tre, I-00146 Rome, Italy
\\
\vskip .1cm
and
\\
CERN, Department of Physics, Theory Division
\\ 
CH-1211 Geneva 23, Switzerland
\\
\vskip .2cm
{\large Davide Meloni}~\footnote{e-mail address: meloni@fis.uniroma3.it}
\\
\vskip .1cm
Dipartimento di Fisica `E.~Amaldi', Universit\`a di Roma Tre
\\ 
INFN, Sezione di Roma Tre, I-00146 Rome, Italy

\end{center}
\vskip 0.7cm
\begin{abstract}
We consider future experiments to detect CP violation in neutrino oscillations and discuss how to test that all asymmetries are indeed described in terms of the single leptonic Jarlskog invariant as predicted in the absence of new physics effects.
\end{abstract}
\end{titlepage}
\setcounter{footnote}{0}
\vskip2truecm

\section{Introduction}

In recent years the experimental study of neutrino oscillations has much contributed to our knowledge of particle physics by establishing non vanishing neutrino masses and by measuring or constraining the corresponding mixing angles. In the near future, within the domain of neutrino oscillations, the most important tasks for experiments, beyond refining the measured values that are already known,  are the measurement of the mixing angle $\theta_{13}$, which at present is only limited by an upper bound, the determination of the sign  of $\Delta m_{31}^2=m_3^2-m_1^2$, which would solve the existing ambiguity between normal (positive sign) and inverse (negative sign) hierarchy in the neutrino mass spectrum, and finally the observation of CP violation. Once a positive signal for CP violation in neutrino oscillations has been established, the next question will be, like in the analogous case for quarks, whether the observed phenomena are in agreement with the simplest picture where all CP violating quantities in oscillation phenomena are described in terms of only one phase $\delta$ appearing in the PMNS mixing matrix, or equivalently in terms of the Jarlskog invariant J, which in the standard notation for mixing angles is given by:
\be
J= \frac{1}{8}\cos{\theta_{13}}\sin{2\theta_{13}}\sin{2\theta_{23}}\sin{2\theta_{12}}\sin{\delta}.\\
\label{jal}
\ee
As well known, in the mass matrix of 3 Majorana neutrinos two additional CP violating phases also appear. As they do not contribute to CP violation in neutrino oscillations we will not consider them here, but they can have observable effects in other phenomena like neutrino-less double beta decay and leptogenesis. A lot of theoretical effort has been devoted to develop and discuss the strategies for measuring $\delta$ in long baseline experiments with Super-Beams, $\beta$- beams and Neutrino Factories. The  most suitable observables for this purpose are the asymmetries associated with the "golden" channel $\nu_e \leftrightarrow \nu_\mu$ or $\bar{\nu}_e \leftrightarrow \bar{\nu}_\mu$. However other channels are also important to disentangle a set of discrete ambiguities ("degeneracies") in the solutions for $\theta_{13}$ and $\delta$ that arise in the procedure: the "intrinsic" degeneracy due to the functional
dependence of the asymmetry on $\theta_{13}$ and $\delta$, the "sign" degeneracy that occurs if the sign  of $\Delta m_{31}^2$ is unknown and the "octant" degeneracy that depends on the sign of $\theta_{23}-\pi/4$. Each ambiguity is 2-fold, so that a total of 8 solutions are generated if all options are open. The intrinsic degeneracy would remain even if the signs of $\Delta m_{31}^2$ and of $\theta_{23}-\pi/4$ will be known at the time of the measurement. The next important channel, the "silver" channel $\nu_e \leftrightarrow \nu_\tau$ or $\bar{\nu}_e \leftrightarrow \bar{\nu}_\tau$ can be very useful to remove the ambiguity, as well as the $\nu_\mu \leftrightarrow \nu_\tau$ or $\bar{\nu}_\mu \leftrightarrow \bar{\nu}_\tau$. In this respect also the possibility of varying the $L/E_\nu$ ratio of the baseline lenght $L$ and the beam energy $E_\nu$ has been discussed in the literature.
                            
In the present article we address the related problem of testing the PMNS mechanism for CP violation (indicated here as the "standard" framework). We want to study the most suitable tests and discuss the required sensitivity. In order to get a quantitative estimate of the possible effects of new physics we will consider the new physics  contributions to the relevant observables  in a model where the PMNS unitarity is relaxed. This model, denoted as the Minimal Unitarity Violation model (MUV), has been introduced in refs. \cite{Antusch:2006vwa} and further studied in \cite{gavela}, where the present bounds on the new parameters of the model have also been derived. Although this model is certainly not general it is however sufficiently structured to provide an indicative estimate of the possible deviations from the standard description of leptonic CP violation that one may expect\footnote{Previous studies of new sources of CP violation in the context of non-unitary neutrino mixing matrix can be found in \cite{Xing:2007zj} and \cite{Luo:2008vp}.}.  Different models were studied in the literature with new physics interactions at the source and at the detection of neutrinos in oscillation experiments \cite{sou} and also during the propagation in matter \cite{prop}. All these models refer to new interactions, with both diagonal and non diagonal lepton flavour properties, parameterized in terms of effective 4-fermion vertices with couplings left free but constrained by existing data. In the MUV model there are less parameters, so that there are more rigid constraints from the data. The resulting MUV framework is one that reproduces the broad features of the more general models but where the parameters at the source, at detection and in matter are related. As the role of MUV for us is to show that, after imposing the existing experimental bounds, there is still a lot of space for new physics to affect the values of the CP violating observables, the fact that there is even more freedom is not a problem. But the presence of new physics can also have additional implications. For example we will assume in the following that, by the time of the CP violation measurements, some quantities will be known within a reasonable precision, the most important one being the value of $\theta_{13}$. Then, in the presence of new physics,  the results of these measurements could be distorted by the new effects so that they do reproduce the correct values of the relevant quantities. For this purpose, even when the model independence of the results is guaranteed by MUV, it could be invalidated in a more general model. We will discuss this aspect in the following.

The conclusion of our investigation is that the same set of measurements that have been discussed for the determination of $\delta$ and for the removal of degeneracies offer a viable and effective testing ground for the mechanism of CP violation. As already mentioned, we assume that the value of $\theta_{13}$ will be known with sufficient precision in a model independent way by the time of the measurement of CP violation and, as a consequence, we work with $\theta_{13}$ fixed at a set of representative values ($\sin{\theta_{13}}= 0.02,~0.05,~0.10,~0.15$). Within the MUV model the golden channel probabilities are the least affected by new physics, given the powerful constraints arising from the present and forthcoming limits on the $\mu \rightarrow e \gamma$ decay, and can be used for an essentially model independent determination of $\sin{\theta_{13}}$ and also of $\delta$.  If the measured asymmetry is compatible with a discrete set of values of $\delta$ (if not compatible one directly obtains clear evidence for some exotic mechanism) then the silver channel and other possible measurements offer the possibility to check the corresponding predictions of the standard picture. For discussing this comparison we eliminate $\delta$  and express one CP violating quantity in terms of another. For example, a CP violating asymmetry for the silver channel can be given in terms of the measured value of the golden channel asymmetry, or the value of one asymmetry can be predicted in terms of the same asymmetry at a different $L/E_\nu$ value. We show that the present constraints on the MUV parameters still allow a considerable space for results at sizable variance with respect to the standard expectation, so that meaningful tests can be obtained with realistic measurements.
For this to be realised the experimental uncertainty associated with the CP observables in the standard model must be smaller than the deviations induced by the new physics parameters. To check that this is indeed the case, we estimate the errors on the various asymmetries at one particular experimental setup (a neutrino factory with $L=1500$ Km) showing that, even under reasonable assumptions on efficiencies, backgrounds and systematic errors, one can reasonably expect to incisively probe the presence of new physics.
We also show that the degeneracies do not prevent the possibility of testing the standard CP mechanism, at least for not too small values of $\theta_{13}$.

This paper is organised as follows. In Sect. 2 we briefly review the standard formalism. In Sect. 3 we summarise the MUV model and extend the formulae for probabilities and asymmetries to include the new physics effects. In Sect.4 we discuss the set of possible future facilities that we will consider in our analysis. Sect. 5 contains our main results on the way of testing the standard picture of CP violation. Finally in Sect. 6 we summarise our conclusion. 
%
%%%%%%%%%%%%%%%%%%%%%%%% SECTIONS:    %%%%%%%%%%%%%%%%%%%%%%%%%%%%%%
%
\section{Transition probabilities and asymmetries in the PMNS framework}
\label{SM}
In this section we briefly recollect the formalism used to describe standard
neutrino oscillations in matter and in vacuum. The lepton mixing is described by the
$3\times 3$ unitary matrix $U_{PMNS}$ \cite{Pontecorvo:1957cp} which, in its
standard parameterization, has the following form \cite{Chau:1984fp}, \cite {rev}:

\begin{eqnarray}
\label{Eq:upmns}
 U_{PMNS} = \left(
  \begin{array}{ccc}
  c_{12}c_{13} &
  s_{12}c_{13} & s_{13}e^{-I\delta}\\
  -c_{23}s_{12}-s_{13}s_{23}c_{12}e^{I\delta} &
  c_{23}c_{12}-s_{13}s_{23}s_{12}e^{I\delta}  &
  s_{23}c_{13}\\
  s_{23}s_{12}-s_{13}c_{23}c_{12}e^{I\delta} &
  -s_{23}c_{12}-s_{13}c_{23}s_{12}e^{I\delta} &
  c_{23}c_{13}
  \end{array}
  \right) 
\end{eqnarray}
where $s_{ij}$ and $c_{ij}$ are short-hand notations for 
$\sin \theta_{ij}$ and $\cos \theta_{ij}$, respectively
whereas $\delta$ is the (Dirac) CP violating phase\footnote{Possible Majorana phases are not explicitly considered here because they do not manifest themselves in oscillation experiments.}.
Neutrino oscillation probabilities are obtained by solving the Schr\"odinger equation for flavour eigenstates with the effective hamiltonian given by:

\begin{equation}
\label{eq:fullham}
	H = \frac{1}{2E} U_{PMNS} {\rm diag}(0,\Delta m_{21}^2,\Delta m_{31}^2) U_{PMNS}^\dagger
		+ {\rm diag}(V,0,0)
\end{equation}
where $\Delta m_{ij}^2 =
m_i^2-m_j^2$, and ${\rm diag}(V,0,0)$ is the potential felt by neutrinos in their propagation in matter. $V$ is given by:
\bea
\nn V=\sqrt{2}\,G_F\,N_e
\eea
with $G_F$ being the Fermi constant and $N_e$ the ambient electron number density
\cite{Wolfenstein:1977ue}.
Expressions of the transition probabilities $P(\nu_\alpha \to  \nu_\beta)\equiv P_{\alpha\beta}$ 
($P(\bar \nu_\alpha \to \bar  \nu_\beta)\equiv P_{\bar \alpha \bar \beta}$)
in matter can be found in a number of papers (see, e.g., \cite{Akhmedov:2004ny}-\cite{Kimura:2002wd}). All our numerical results in Sect.5 have been obtained by numerically solving the coupled Schroedinger equations involving the appropriate Hamiltonians (for the SM and MUV cases). On the other hand, for the sake of simplicity, the analytical formulae, given in the present section as well as in the next one, have been computed using some useful approximations, motivated by the underlying physics.

At present neutrino oscillation results for the mixing angles \cite{res}, within 1-$\sigma$, are consistent with the so-called $tri-bimaximal$ mixing (TBM) matrix \cite{hps} 
\begin{eqnarray}
U _{PMNS}\approx
\left( \begin{array}{rrr}
\sqrt{\frac{2}{3}}  & \frac{1}{\sqrt{3}} & 0 \\
-\frac{1}{\sqrt{6}}  & \frac{1}{\sqrt{3}} & \frac{1}{\sqrt{2}} \\
\frac{1}{\sqrt{6}}  & -\frac{1}{\sqrt{3}} & \frac{1}{\sqrt{2}}
\end{array}
\right)
\label{MNS0}
\end{eqnarray}
Following \cite{King:2007pr}, we parametrize the deviation from tri-bimaximal mixing by introducing three small parameters $r,s,a$ in such a way that:
\be
s_{13} = \frac{r}{\sqrt{2}}, \ \ s_{12} = \frac{1}{\sqrt{3}}(1+s),
\ \ s_{23} = \frac{1}{\sqrt{2}}(1+a).
\label{deviations}
\ee
The results of global fits on neutrino data 
(\cite{GonzalezGarcia:2007ib},\cite{Strumia:2006db}) imply that these parameters can be at most of $\mathcal O$ (10\%). One can then expand all probabilities and asymmetries in $r,s,a$ around the TBM matrix. We can further simplify the analytical expressions by expanding in the variable $\xi=\Delta m_{21}^2/|\Delta m_{31}^2|\sim 0.03$ and working in the small $\Delta_{21}$ limit, where
$\Delta_{21} = \Delta m_{21}^2 L/ 4E_\nu \ll 1$.
In the following we quote the relevant transition probabilities and the related CP asymmetries up to second order in those small parameters. Note that only r appears in the asymmetries up to second order while a and s appear among the dominant terms only in $P_{\mu \mu}$ (eq. \ref{esse}, \ref{ese}, \ref{essse}) and in $P_{\mu \tau}$ (eq. \ref{aa}). 

First, we consider the {\it golden} channel $P_{e \mu}$ \cite{Cervera:2000kp}:
\bea
\nn
P_{e \mu}&=&r^2\,\frac{\sin^2 \left(A-1 \right) \,\Delta_{31}}{( A-1)^2} + 
  \frac{4}{3}\,r\,\xi \,\cos (\delta  - \Delta_{31})\,\sin (A-1) \,\Delta_{31}\,
     \frac{\sin (A\,\Delta_{31})}{A(A-1)} + \\ &&\frac{4}{9}\,\xi^2\,\frac{\sin^2 (A\,\Delta_{31})}{A^2} 
\label{Pemu}
\eea
where $A=VL/2 \Delta_{31}$. The CP conjugate channel 
$P_{\bar e \bar \mu}$ is obtained from $P_{e \mu}$ by changing $\delta \to -\delta$ and 
$A \to -A$. The resulting expression of the asymmetry $A_{e\mu}$ in matter is quite cumbersome; we then limit ourselves to explicitly quote only the vacuum case which shares some of the relevant features of the corresponding quantity in matter:
\bea
\label{Aemu}
A_{e\mu} &=&\frac{12\,r\,\Delta_{21}\,\sin \delta \,\sin^2 \Delta_{31}}
  {4\,\Delta^2_{21} + 9\,r^2\,\sin^2 \Delta_{31} + 
    6\,r\,\Delta_{21} \,\cos \delta\,\sin 2\,\Delta_{31}}
\eea
Notice that the "T-conjugate" asymmetry $A_{\mu e}$ can be easily calculated by assuming that the density profiles are symmetric under interchange of the position of the neutrino source and detector, so that $P_{\alpha\beta}\,(\delta, A)=P_{\beta\alpha}\,(-\delta, A)$, which in turn implies $A_{\mu e}\,(\delta, A)=A_{e\mu}\,(-\delta,A)$.  Thus, the results for $P_{\mu e}$ and  $A_{\mu e}$ can be directly obtained from  eqs.(\ref{Pemu}-\ref{Aemu}).

In the approximations adopted in this section, the {\it silver} channel $P_{e \tau}$ \cite{Donini:2002rm} has an expression similar to eq.(\ref{Pemu}), the only difference being a sign change in the second term where the CP violating phase appears. This implies:
\bea
\label{Aetau}
A_{e\tau} &=&\frac{12\,r\,\Delta_{21} \,\sin \delta \,\sin^2 \Delta_{31}}
  {-4\,\Delta^2_{21} - 9\,r^2\,\sin^2 \Delta_{31} + 
    6\,r\,\Delta_{21} \,\cos \delta\,\sin 2\,\Delta_{31}}
\eea 

Finally, we quote $A_{\mu \tau}$:

\bea
\label{Amutau}
A_{\mu\tau} &=& \frac{4}{3}\, r\, \Delta_{21} \sin \delta.
\eea 
This simple expression is a consequence of the zeroth-order contribution of  $P_{\mu \tau}$ 
(the expression is given is App.\ref{app:mutau}), which is absent in $P_{e \mu,\tau}$.

Notice that all the asymmetries in vacuum are suppressed by the small quantities $\Delta_{21}$ and $\theta_{13}$; however, since the denominators in eqs.(\ref{Aemu}-\ref{Aetau}) are also suppressed, a partial cancellation is at work and, in particular, one generically expects $A_{e \mu},A_{e\tau}\gg A_{\mu \tau}$.

Matter effects play an important role in modifying the behaviour of the asymmetries as function of the CP phase $\delta$ (see, e.g., \cite{Donini:1999jc} and \cite{Dick:1999ed}). In fact, while in the vacuum case $A_{\alpha\beta}=0$ when $\sin \delta=0$, this is no longer true in the more general case because the passage through matter introduces fake CP-violating effects in neutrino propagation. Then, to extract genuine CP violating effects, one often defines the {\it subtracted} asymmetries as  $A^{\rm sub}_{\alpha\beta}(\delta)=A_{\alpha\beta}(\delta)-A_{\alpha\beta}(\delta=0)$. We prefer to deal with more directly measurable quantities, so that we use unsubtracted asymmetries which, for non negligible matter effects, are non vanishing at $\delta=0$.

To conclude this section, we want to mention that also the CP-conserving channel $P_{\mu \mu}$ can be in principle used to extract information on the CP phase\footnote{See app.\ref{app:mumu} for details.} because of the term $r\,\Delta_{21}\cos \delta\sin 2\Delta_{31}$, which shows the same suppression as $A_{\mu\tau}$. In particular, in \cite{Kimura:2006rp} it has been shown that the variation of $P_{\mu \mu}$ due to a change in the value of $\delta$ can be as large as 30\% when $L>2000$ Km and $E_\nu < 2$ GeV.

\section{Summary of the MUV model}
\label{summuv}
As a representative framework for physics beyond the standard model affecting neutrino oscillations we consider the Minimal Unitarity Violation model (MUV) introduced in 
\cite{Antusch:2006vwa} and further studied in \cite{gavela}. This framework is based on the possibility that, although the complete theory is unitary, the effective leptonic mixing matrix $U_{PMNS}$ describing neutrino oscillations at low energy is not unitary, as is the case in a large class of theories addressing the question of neutrino masses.
The deviation from unitarity is parametrized in terms of a Hermitian matrix $\eta$ in such a way that (using the same notation as in \cite{gavela}):
\bea
N=(1+\eta)\,U_{PMNS}
\nn
\eea
where $N$ is the non-unitary leptonic mixing matrix relating the interaction and mass  eigenstates, $\nu_\alpha=N_{\alpha i}\, \nu_{i}$.
Being a hermitian matrix, $\eta$ contains 9 new parameters, for example 6 moduli and three phases. In particular, the phases are new sources of CP violation in the leptonic sector. The moduli of the $\eta$ matrix elements are bounded at the level of $\mathcal O$(1\%) (or smaller) from universality tests, rare lepton decays, the invisible width of the Z boson and neutrino oscillation data \cite{Antusch:2006vwa}:
\bea
|\eta_{ee}| \sim |\eta_{\mu\mu}|&\sim& |\eta_{\tau\tau}| < 2.5 \times 10^{-3} \nn \\
|\eta_{e\mu}|&<& 3.6 \times 10^{-5} \nn \\
|\eta_{e\tau}|&<& 8 \times 10^{-3} \nn \\
|\eta_{\mu\tau}|&<& 4.9 \times 10^{-3}. 
\label{boundMUV}
\eea
On the other hand the phases, as for the standard CP phase in $U_{PMNS}$, are completely unbounded.  As mentioned in the Introduction, in models where new physics contributions are introduced
at all stages in the oscillation process, starting from the source of neutrinos and then in the interaction with matter during the propagation and finally in the detection process,  there is space for a wider range of effects. In fact the new physics allowed by MUV is more limited than in general. But our main purpose in showing the MUV predictions is to prove that, in each specific case, there can be sufficiently large new physics effects to be detected. On the other hand, while we have specified the set of present constraints, it must be kept in mind that more restrictive limits could be obtained by the time when the CP violation experiments will be performed.

\subsection{Transition probabilities and asymmetries in MUV}

In the MUV case, since our aim is to propose consistency checks of the leptonic CP sector of the
standard model based on the currently planned experimental
options, we present analytic transition probabilities and asymmetries with the
same approximations used in Sect.\ref{SM}.
Furthermore, we also expand up to second order in the small
$\eta_{\alpha\beta}$'s even if, from the limits given in eq.(\ref{boundMUV}),
the retained $\eta$ terms could be smaller than the neglected terms in the $r,s$
and $a$ expansion (defined in eq.(\ref{deviations})).
The asymmetries including the MUV corrections are derived using the procedure described in Appendix A of  \cite{gavela}.
Based on the discussion of the previous section, in the following we only quote
the vacuum results, which are enough to understand the
role of the MUV parameters compared to the standard model results. 

In the MUV framework, the probability $P_{e\mu}$ mainly depends on the
parameters $\eta_{e\mu}$ and $\eta_{e\tau}$ and their related phases. The
detailed expression is given in App.\ref{app:emu}; here we only retain terms which are linear in the small $\eta_{\alpha\beta}'s$ and do
not depend on the product $\eta_{\alpha\beta}\,\Delta_{21}$, obtaining:

\bea
\label{eq:pemumuv}
\nn
P_{e\mu} &=& P_{e\mu}^{\rm SM}+\eta_{e\mu}r\,\sin\Delta_{31}\left[3\sin(\delta-\Delta_{31}-\delta_{e\mu})+
\sin(\delta+\Delta_{31}-\delta_{e\mu})\right] \\ &&
+ 2\eta_{e\tau} r\,\cos(\delta-\delta_{e\tau})\sin^2\Delta_{31}
\eea 
where by $P^{\rm SM}_{\alpha\beta}$ we denote the SM results.
The asymmetry $A_{e\mu}$ obtained from  eq.(\ref{eq:pemumuv}) reads:
\bea
\\ \nn
\hspace{-0.4cm}
A_{e\mu} &=&\frac{12\,r\,\Delta_{21}\,\sin\delta\,\sin^2 \Delta_{31} +
  18\,r\,\eta_{e\mu}\,\sin 2\Delta_{31}\,\sin(\delta - \delta_{e\mu })}
{4\,\Delta_{21}^2+9 r^2\sin^2\Delta_{31}+18 r \sin^2\Delta_{31}[\eta_{e\tau}\cos(\delta-\delta_{e\tau})-
\eta_{e\mu}\cos(\delta-\delta_{e\mu})]+6 r \Delta_{21}\cos \delta \sin 2\Delta_{31}}
\eea
Similarly to the standard model case, the modified $A_{\mu e}$ can be obtained from $A_{e \mu}$ changing the sign of all the phases. We see that a new physics term proportional to $\eta_{e\mu}$ appears in the numerator with a different dependence on $L/E_\nu$; however,
in the MUV model, due to the strong bounds on the parameter
$\eta_{e\mu}$, we can safely neglect this term, containing a new CP violation effect.
Note also that
the new terms in the denominator show
a  dependence on $L/E_\nu$ which is the same as the standard terms so that  their contribution (and in particular that due to
$\eta_{e\tau}$) cannot compete against $r$ unless $\theta_{13}$ is very small. As a result the asymmetry $A_{e \mu}$ does not deviate much from the standard result.
It follows that, in the MUV model,  this channel is particularly suitable for an experimental
determination of the $U_{PMNS}$ parameters, as we now briefly discuss. But it must be kept in mind that in a more general model there can certainly be new physics effects in $A_{e \mu}$ leading to larger deviations from the standard picture of CP violation than in MUV. 

In the following analysis we assume that the value of $\theta_{13}$ will be known with sufficient precision by the time that a
meaningful measurement of the CP violating phase $\delta$ can be performed. As we are contemplating the possibility of important new
physics effects, an obvious question is whether $\theta_{13}$ can be determined in a reasonably model independent way.
In the MUV framework, $P_{e\mu}$ and the corresponding asymmetry $A_{e \mu}$ are optimal in this respect as the new physics effects are already
bound to be reasonably small. The planned experimentation at T2K \cite{Itow:2001ee} indeed plans to measure $\theta_{13}$ through this channel.
Alternatively, reactor experiments like
 Double Chooz \cite{Ardellier:2006mn}, would measure $\theta_{13}$ from the diagonal transition probability $P_{ee}$ which,
in the limit of vanishing $\eta_{e\mu}$, is given by:
\bea
P_{ee}&=&  1-2\,\sin^2 \Delta_{31}\,\left[r^2+
\eta_{e\tau}^2+2\,r\,\eta_{e\tau}\cos(\delta-\delta_{e\tau})\right]
-\frac{8}{9} \Delta_{21}^2
\eea
In this channel the new physics effects can indeed be large. Thus a measurement of 
$\theta_{13}$ from reactors only would not guarantee a model independent determination even in the MUV case. But the combination of reactors with T2K would answer
the question on model independence and actually could put bounds on the maximal value allowed for $\eta_{e\tau}$ in the MUV model. In a more general context than MUV different measurements are even more necessary in order to put limits on the model dependence of the measured value of $\theta_{13}$. In the following we will assume that the value of $\theta_{13}$ is known, but it is well possible that the issue of a truly model independent determination of $\theta_{13}$ will only  be solved by the same generation of experiments that will tackle the CP violation problem.

Important corrections to the SM results may appear in $P_{e\tau}$ which, for $\eta_{e\mu}=0$, is given by:
\bea
\nn
P_{e\tau} &=& P^{\rm SM}_{e\tau} + \eta_{e\tau}\,\left\{r \sin\Delta_{31}\left[3\sin (\delta-\delta_{e\tau}-\Delta_{31})+\sin (\delta-\delta_{e\tau}+\Delta_{31})\right]-\right.\\
&&\left.\frac{2}{3}\Delta_{31}\,\xi\,[\sin (2\Delta_{31}-\delta_{e\tau})-3\sin \delta_{e\tau}]\right\}
+\frac{1}{2}\,\eta^2_{e\tau}\,(5+3\cos 2\Delta_{31})
\eea

The new source of CP violation is only linearly suppressed in the small $\eta_{e\tau}$ parameter; in particular, the first term in curly brackets can, in principle, compete with the standard term $\frac{4}{3}\,r\,\xi \,\cos (\delta  - \Delta_{31})\,\Delta_{31} \,\sin \Delta_{31}$ because it is not suppressed by the small quantity $\Delta_{21}$. Note also that the quadratic term corresponds to the zero-distance effect discussed in \cite{Antusch:2006vwa}. From $P_{e\tau}$, the asymmetry $A_{e\tau}$ reads:

\bea
\\ \nn
\hspace{-0.4cm}
A_{e\tau} &=&\frac{-8\,r \Delta_{21}\,\sin\delta\,\sin^2 \Delta_{31} +
  4\,\eta_{e\tau}\,\left[3 r \sin 2\Delta_{31}\,\sin(\delta - \delta_{e\tau })+\Delta_{21} \sin \delta_{e\tau} (3+\cos 2\Delta_{31})\right]}
{8/3 \Delta_{21}^2+6\,r^2\sin^2\Delta_{31}-4\eta_{e\tau}\left[\Delta_{21}\sin 2\Delta_{31}+3r\cos(\delta-\delta_{e\tau})\sin^2\Delta_{31}\right]-4r \Delta_{21}\cos\delta\sin 2\Delta_{31}}
\eea
where, for simplicity,  we have disregarded  terms proportional to $\eta_{e\tau}^2$ and $\eta_{e\tau}\Delta_{21}$.
It is clear that, whenever $\Delta_{21} \sim \eta_{e\tau}$, the asymmetry can be quite different with respect to the SM result.  Then  $A_{e\tau}$ can be a good CP-violating observable to perform consistency checks of the leptonic CP sector of the SM. 

A quite different case is that related to $A_{\mu\tau}$; as illustrated in eq.(\ref{Amutau}), in the SM $A_{\mu\tau}$ is somewhat limited to small values, around $10^{-3}$ for the maximum allowed $\theta_{13}$ and $\delta$. Thus it would be much more easy to see any deviation from standard results if new physics would affect  $\nu_\mu \to \nu_\tau$ transitions \cite{gavela}. This is indeed the case since the main correction to the SM expression of $P_{\mu\tau}$ is linear in the new parameter $\eta_{\mu\tau}$:
\bea
\label{eq:pmutau}
P_{\mu\tau}=P^{SM}_{\mu\tau}-2 \,\eta_{\mu\tau}\sin 2 \Delta_{31} \sin \delta_{\mu\tau}
\eea
and, consequently:
\be
\label{eq:Amutau}
A_{\mu\tau}=A^{SM}_{\mu\tau}-4 \,\eta_{\mu\tau} \cot\Delta_{31} \sin \delta_{\mu\tau}
\ee
where the SM term is suppressed by at least one power of $\Delta_{21}$. The new physics term easily overwhelms the standard one and $A_{\mu\tau}$ can reach values as large as $\mathcal O$ $(10^{-1})$.

Let us finally comment on $P_{\mu\mu}$ in the MUV scheme. The correction to the
SM results is quadratic in the small parameters of our
perturbative expansions and only depends on $\eta_{\mu\tau}$ (but
not in $\eta_{\mu\mu}$):

\bea
P_{\mu\mu}=P^{SM}_{\mu\mu} +4\,\eta_{\mu\tau} \cos\delta_{\mu\tau}\,\sin^2 \Delta_{31}\,(2\,a+\eta_{\mu\tau}\,\cos \delta_{\mu\tau})
\label{esse}
\eea

\begin{figure}[h!]
\centering
\includegraphics[width=0.75\linewidth]{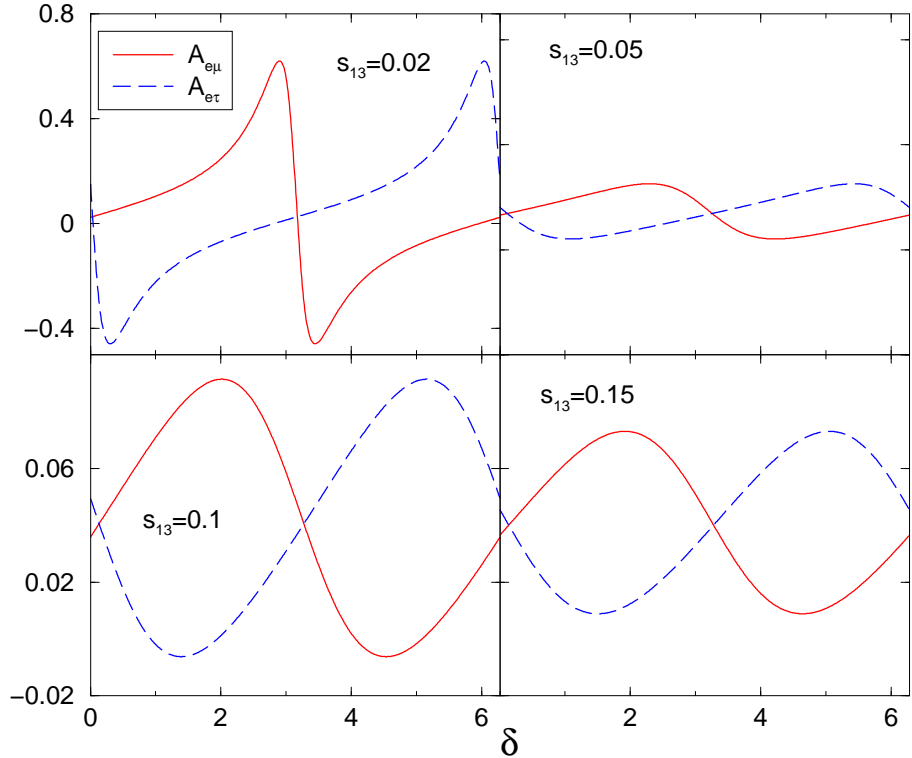}
\caption{\label{fig:asvsdelta} \it The standard model asymmetries $A_{e\mu}$,  $A_{e\tau}$  for different values of 
$\sin{\theta_{13}}= s_{13}=0.02,~0.05,~0.10,~0.15$ as a function of the phase $\delta$. We have assumed that the sign of $\Delta_{31}$ is positive (normal hierarchy), that $\theta_{23}$ is maximal and $\sin^2{\theta_{12}}=1/3$.
We have also fixed the mass differences to $\Delta m^2_{21}=8\times 10^{-5}$ eV$^2$   and $\Delta m^2_{31}=2.4\times 10^{-3}$ eV$^2$.
Neutrino energy and baseline distance $L$ are fixed to  $E=30$ GeV and $L=1500$ Km, respectively.}
\end{figure}
                                                           
\begin{figure}[h!]
\centering
\includegraphics[width=0.75\linewidth]{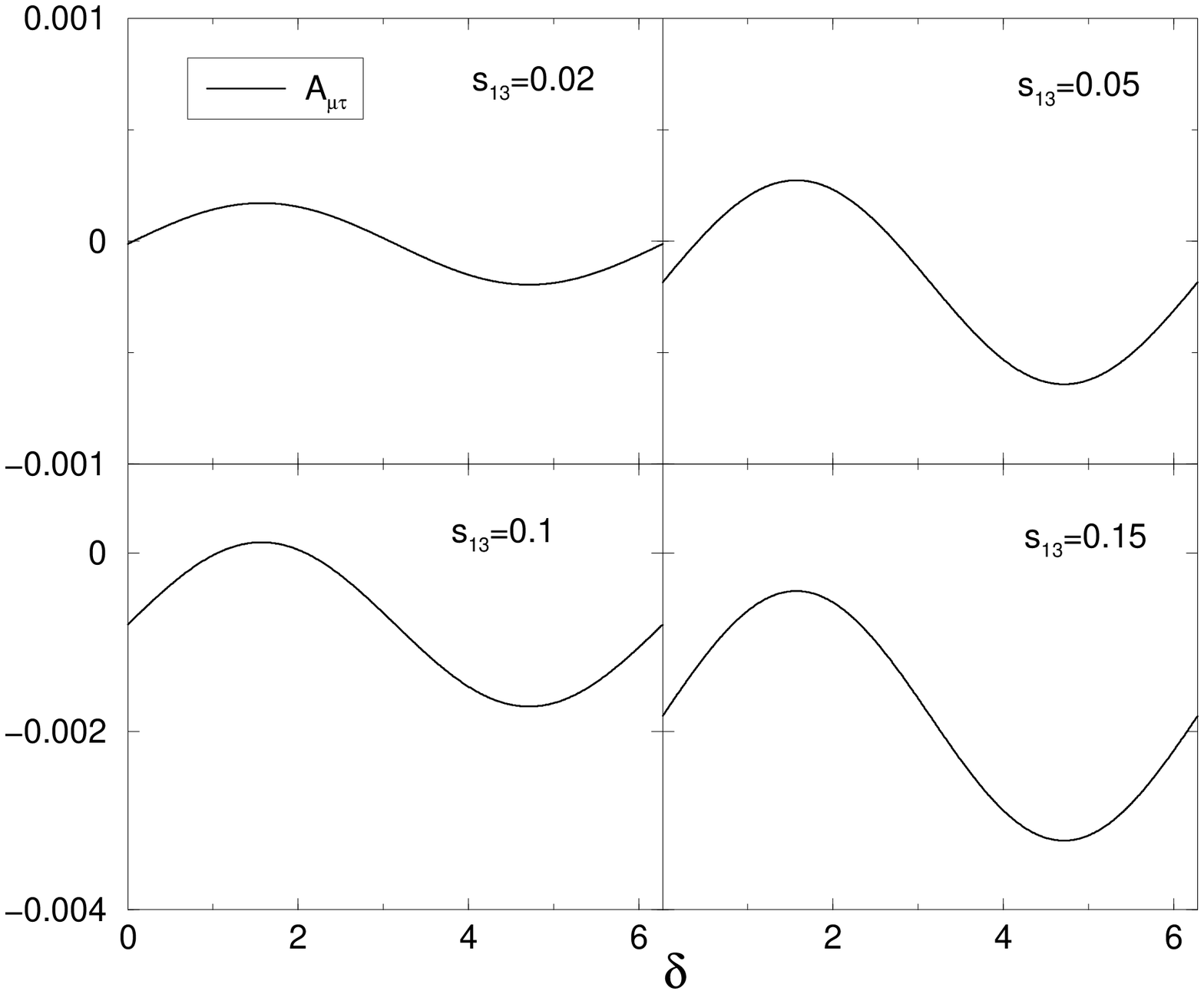}
\caption{\label{fig:asvsdelta2} \it The standard model asymmetry $A_{\mu\tau}$ for different values of 
$\sin{\theta_{13}}= s_{13}=0.02,~0.05,~0.10,~0.15$ as a function of the phase $\delta$. All other parameters are fixed as in the previous figure. Neutrino energy and baseline distance $L$ are fixed to  $E=35$ GeV and $L=1500$ Km, respectively.}
\end{figure}

\section{A list of possible future experiments}
\label{catalogo}

In this section we summarize the experimental facilities we consider in our analysis. In recent years there has been a vast amount of theoretical work on the performances of several setups  aimed at studying some of the still unknown neutrino parameters and properties, like the value of $\theta_{13}$, the existence of leptonic CP-violation, the type of mass hierarchy and the precise value of $\theta_{23}$. In particular, to assess the problem of CP violation, it has been pointed out that the zeroth-order requirement is to have very intense neutrino fluxes and good detector performances: this automatically aims at second-generation neutrino experiments, like Super-Beams, $\beta$-beams and Neutrino Factories. Since the measurement of the standard CP phase $\delta$ is directly connected with the solution of the problem of the degeneracies \cite{Burguet-Castell:2001ez}-\cite{Barger:2001yr}, one needs to combine different transition channels (golden, silver and also $\nu_\mu \to \nu_\tau$ channels), or to exploit several $L/E_\nu$ regimes (which can be accomplished either using two detectors located at different baselines or by varying the neutrino energy). Also, a combination of these facilities among themselves and/or in association with atmospheric neutrino data can be quite useful for constraining $\delta$. For a review of these possibilities, see \cite{Bandyopadhyay:2007kx}.
For our purposes, it is sufficient to select some of the most promising future experiments for the measurement of $\delta$ and probe their ability to perform the consistency checks we aim to. The extension of the analysis to other facilities and combinations is straightforward.
\bi
\item  Super-beams

These are conventional neutrino beams from $\pi$ decay (except for their large intensity). Since the flux is predominantly $\nu_\mu$, the  $\nu_\mu \to \nu_\mu$ and  $\nu_\mu \to \nu_e$ transition channels are available (the $\nu_\mu \to \nu_\tau$ transition can only be probed using high energy neutrino fluxes having sufficient energy to produce $\tau$ leptons).
The second generation of upgraded Super-Beams can run with both neutrino helicities, thus having sensitivity to CP violation; we refer to two different options: the upgraded version of T2K \cite{Itow:2001ee}, T2HK, with $L=295$ Km and $\langle E_\nu \rangle \sim 0.75$ GeV  and the CERN Super-Beam project SPL \cite{SPL}
with $\langle E_\nu \rangle \sim 0.3$ and $L=130$ (Frejus) or $L=732$ (Gran Sasso);

\item  $\beta$-Beams

These are intense $\nu_e$ and $\bar \nu_e$ fluxes produced from boosted radioactive ion decays \cite{Zucchelli:2002sa}; then it is possible to study $\nu_e \to \nu_e$, $\nu_e \to \nu_\mu$ 
(and probably $\nu_e \to \nu_\tau$ if very high-energy options will be availabe) and their CP-conjugate channels. Depending on the choice of decaying ions and boost $\gamma$ factors, several possibilities could be exploited. Here we refer to the standard {\it low-energy}  $\beta$-beam, with $\langle E_\nu \rangle \sim 0.4$ GeV and $L=130$ Km (LE$\beta$B) \cite{BurguetCastell:2005pa} and {\it high-energy} $\beta$-beam (HE$\beta$B)
with $\langle E_\nu \rangle \sim 1$ GeV and $L\sim 732$ Km
\cite{BurguetCastell:2005pa}-\cite{Agarwalla:2008ti}.

\item  $\nu$ Factories

In a neutrino factory, neutrinos are produced via the decay of muons, which are accumulated in a storage ring and accelerated to the desired energy. The resulting neutrino flux is analitically calculable from three-body decays kinematics and strongly depends on the parent muon energy. Since two different types of neutrinos are produced for fixed muon charge, there are in principle six different channels that can be simultaneously probed, assuming that the appropriate detector technologies are available. Then, in principle, the three asymmetries $A_{e\mu}$, $A_{e\tau}$ and $A_{\mu\tau}$ can be investigated, as well as the $\nu_\mu$ disappearance. 
If we concentrate on the sensitivity to CP violation, the performance on a
neutrino factory especially depends on the values of $\theta_{13}$.
For smaller values,
the main uncertainties in the
$\delta$ extraction come from the degeneracy problem and a good performance is
reached for $E_\mu > 30$ GeV and $L\in[3000-5000]$ Km \cite{Huber:2006wb}
(especially if in combination with another experiment running at the "magic"
baseline $L\sim 7500$ Km \cite{Huber:2003ak}) whereas  for larger values
matter effects are the main source of uncertainty and
a smaller baseline is preferable\footnote{Notice also that a
low-energy neutrino factory
\cite{Geer:2007kn} ($E_\mu =4.12$ GeV) seems to have a good performance to CP violation for intermediate 
$\theta_{13} \sim 3\,^o$.}.
Here we consider a neutrino factory with $E_\mu=50$ GeV, for which  $\langle E_{\nu_e,\bar\nu_e} \rangle=30$ GeV and
$\langle E_{\nu_\mu,\bar\nu_\mu} \rangle=35$ GeV.
\ei
In Tab.\ref{tab:comp}, we summarize the setups considered in our analysis.

\begin{table} 
\begin{center}
\begin{tabular}{||c|c|c|c|c||}
\hline
\hline
 &  \it available channels & \it experiments & $L$ (km) & $\langle E_\nu \rangle$ (GeV)\\ 
\hline
 & $\nu_\mu \to \nu_\mu$ & T2HK & 295 & 0.75\\ 
 \it Super-Beams &  $\nu_\mu \to \nu_e$ & SPL & 130-732 & 0.3\\
\hline
& $\nu_e \to \nu_e$  & LE$\beta$B & 130&0.4\\
\it $\beta$-beams & $\nu_e \to \nu_\mu$  &   &  &\\ 
&($\nu_e \to \nu_\tau$) & HE$\beta$B & 732 & 1\\
\hline
& $\nu_e \to \nu_{e,\mu,\tau}$ & NF@4000 & 4000 &  30($\nu_e,\bar
\nu_e$)/35($\nu_\mu,\bar \nu_\mu$)\\
\it Neutrino Factories & $\bar \nu_\mu \to \bar \nu_{e,\mu,\tau}$ &  &  & \\ 
& + CP-conjugates &NF@1500 &1500 &30($\nu_e,\bar
\nu_e$)/35($\nu_\mu,\bar \nu_\mu$)\\
\hline
\hline
\end{tabular}
\end{center}
\caption{\it \label{tab:comp}Summary of the relevant channels and parameters of the future neutrino facilities considered in this work. In the second column we list all the possible available channels at a given facility, assuming that the adequate detector technology will be available.}
\end{table}

\begin{figure} [h!]
\centerline{\epsfig{figure=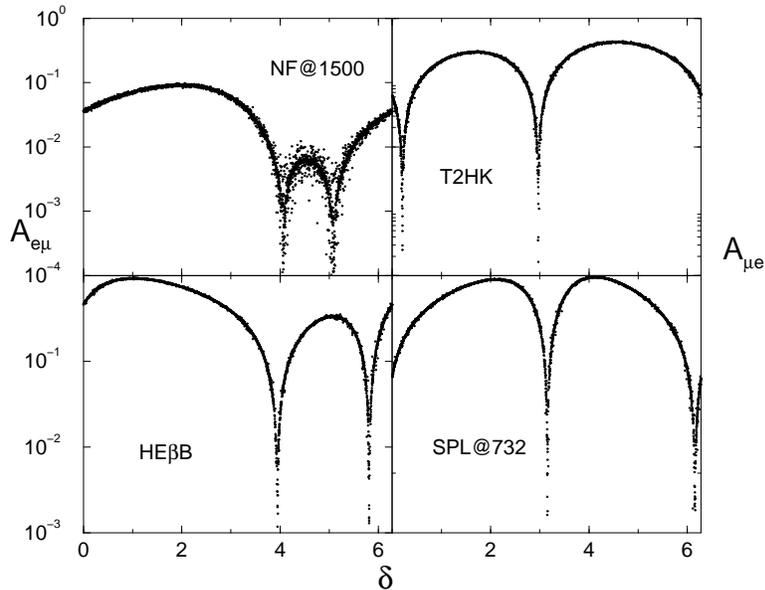,angle=0,width=10cm}}
\caption {\it \it Scatter plots, for the absolute values of the asymmetries $A_{e\mu}$ and $A_{\mu e}$ of the domains spanned when the MUV parameters are moved in their allowed ranges. The angle $\theta_{13}$ is fixed to
$\sin{\theta_{13}}=0.1$. The values of the other parameters are as in the previous figures. Energies and baselines match the values quoted in Tab.\ref{tab:comp}. As explained in the text, in the MUV model the deviations from new physics are particularly small for these asymmetries.
\label{fig:Aem}}
\end{figure}

\begin{figure} [h!]
\centerline{\epsfig{figure=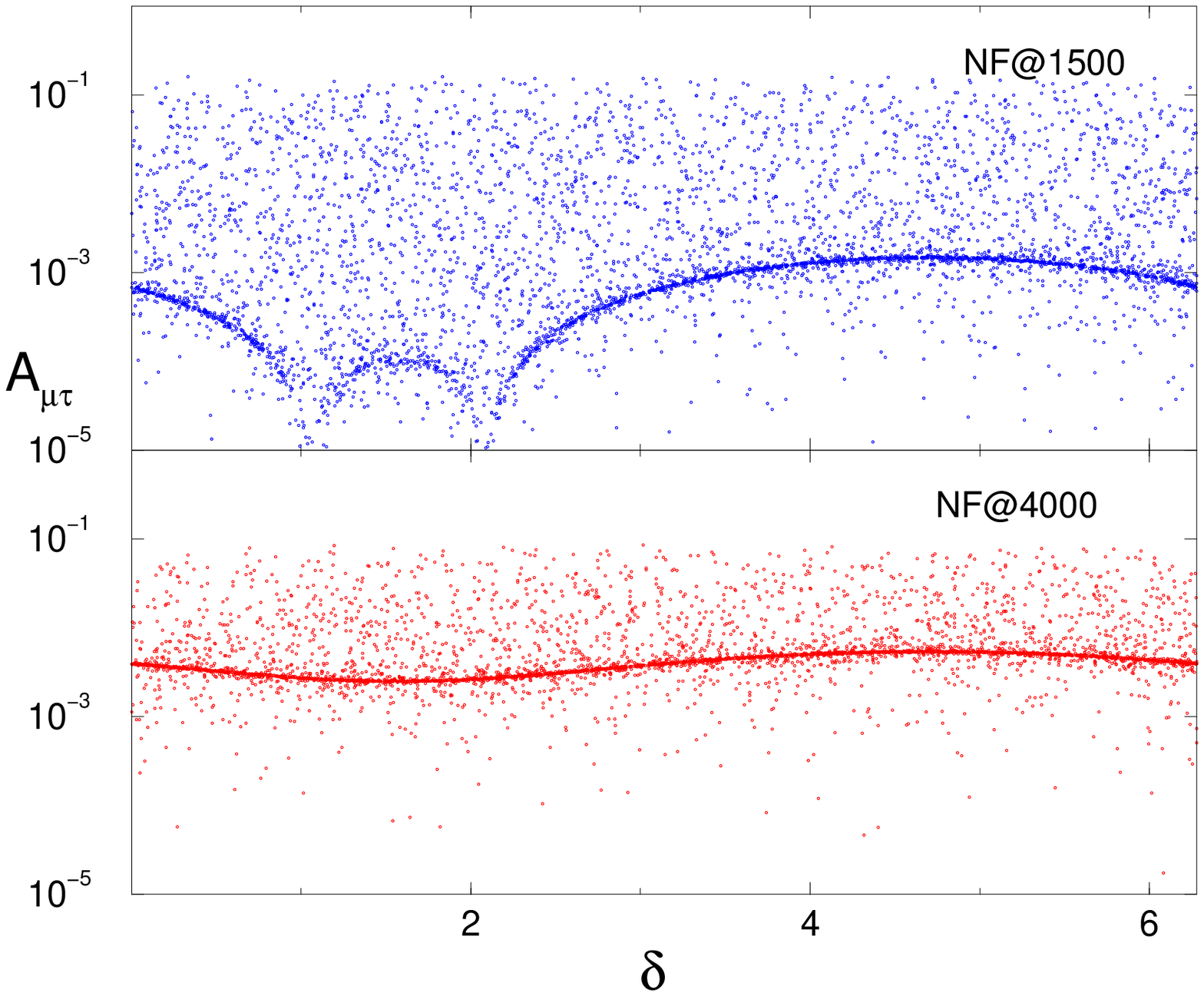,angle=0,width=10cm}}
\caption {\it Scatter plots, for the absolute values of the asymmetry $A_{\mu\tau}$ of the domains spanned when the MUV
parameters are moved in their allowed ranges. The angle $\theta_{13}$ is fixed to
$\sin{\theta_{13}}=0.1$. The values of the other parameters are as in the previous figures. For any of the two facilities, we consider $E_{\nu}=35$ GeV.
\label{fig:Amt}}
\end{figure}

\begin{figure} [h!]
\centerline{\epsfig{figure=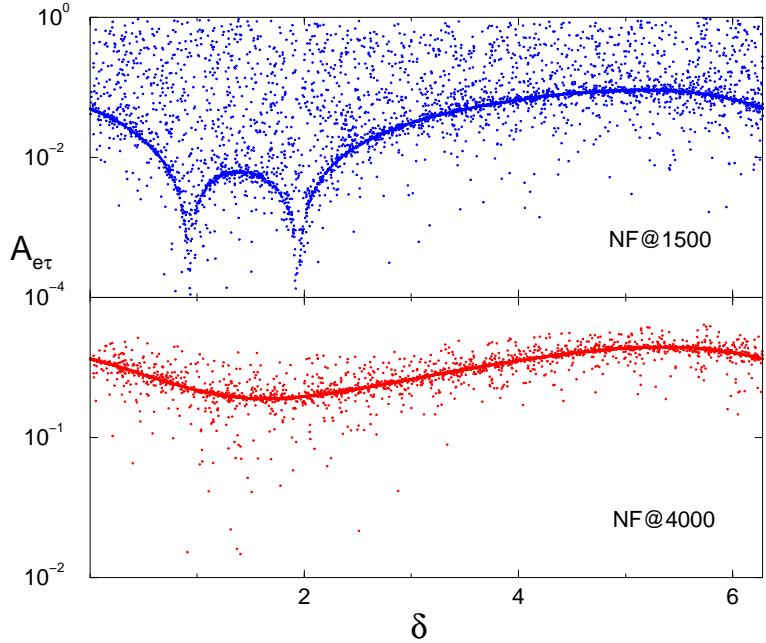,angle=0,width=10cm}}
\caption {\it Scatter plots, for the absolute values of the asymmetry $A_{e\tau}$ of the domains spanned when the MUV
parameters are moved in their allowed ranges. The angle $\theta_{13}$ is fixed to
$\sin{\theta_{13}}=0.1$. The values of the other parameters are as in the previous figures. For any of the two facilities, we consider $E_{\nu}=30$ GeV.
\label{fig:Aet}}
\end{figure}

\section{Testing the CP violation mechanism}

 We start by recalling the dependence of the $A_{e\mu}$,  $A_{e\tau}$ and $A_{\mu\tau}$ asymmetries on the PMNS 
phase $\delta$ in one typical long baseline experiment with $L=1500$ Km and $E_\nu=30$ GeV (for $A_{e\mu}$ and $A_{e\tau}$) 
or $E_\nu=35$ GeV (for $A_{\mu\tau}$). In Figs.\ref{fig:asvsdelta} and \ref{fig:asvsdelta2} we plot the 
asymmetries vs $\delta$ for fixed values of $\theta_{13}$, $\sin{\theta_{13}}= 0.02,~0.05,~0.10,~0.15$, assuming that the sign of $\Delta_{31}$ is positive (normal hierarchy), that $\theta_{23}$ is maximal and $\sin^2{\theta_{12}}=1/3$.
We also fixed the mass differences to $\Delta m^2_{21}=8\times 10^{-5}$ eV$^2$   and $\Delta m^2_{31}=2.4\times 10^{-3}$ eV$^2$.
The curves refer to the conventional picture (all MUV parameters set to vanish).  We see that, at fixed $L/E_\nu$, for a given value of each asymmetry there is in general a two-fold ambiguity in the corresponding value of $\delta$.

\begin{figure}[h!]
\centering
\includegraphics[width=0.49\linewidth]{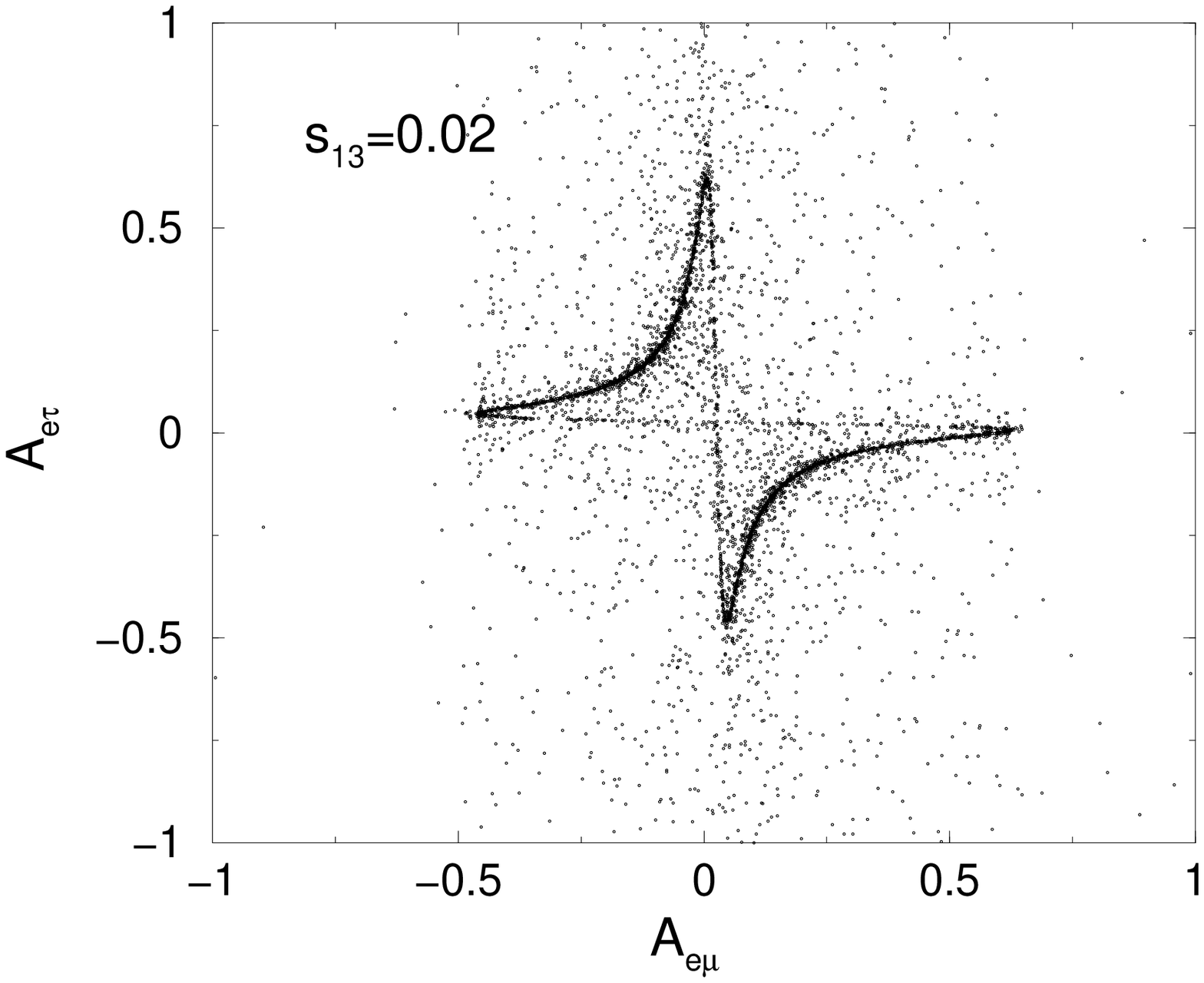} \includegraphics[width=0.49\linewidth]{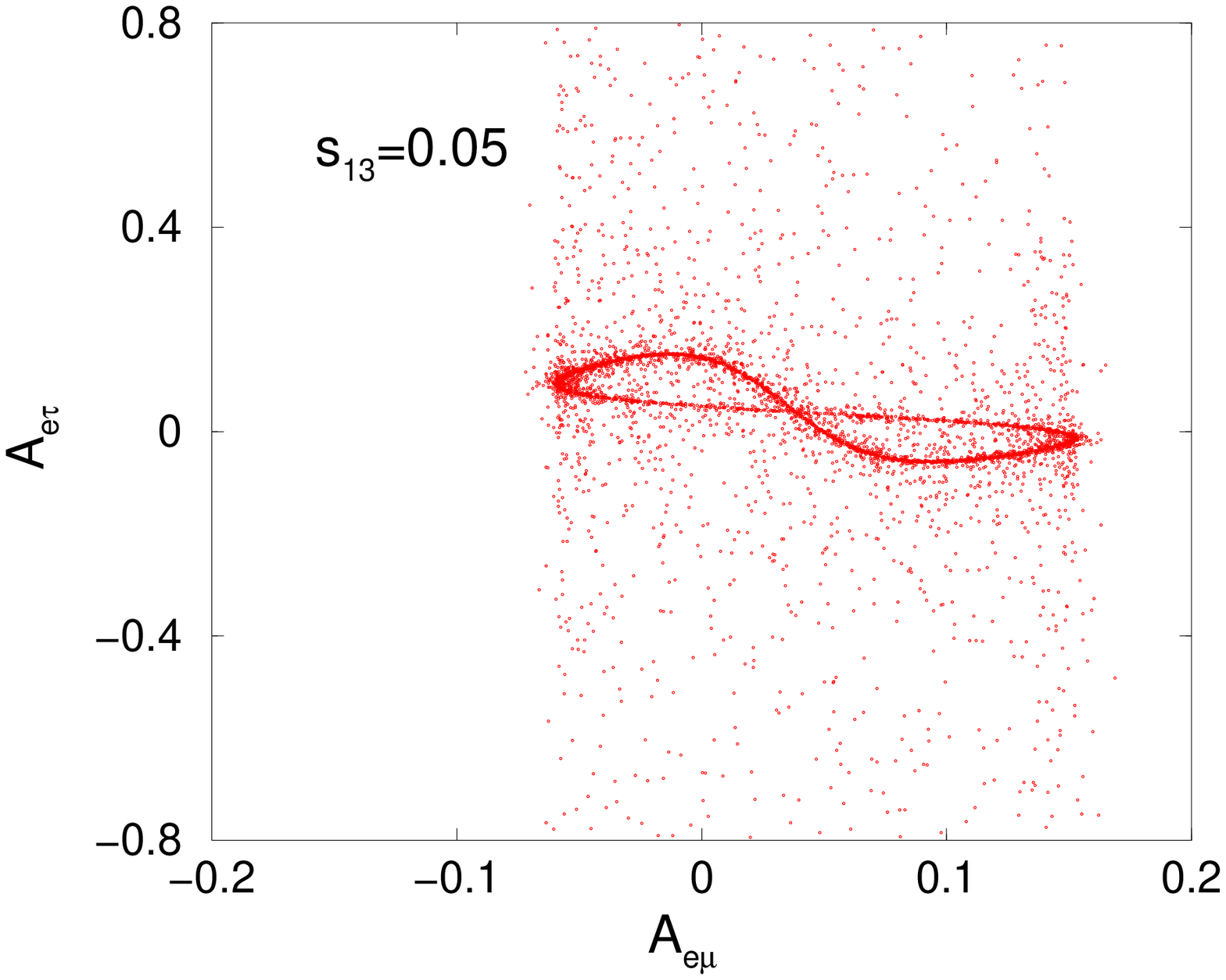}\\
\includegraphics[width=0.49\linewidth]{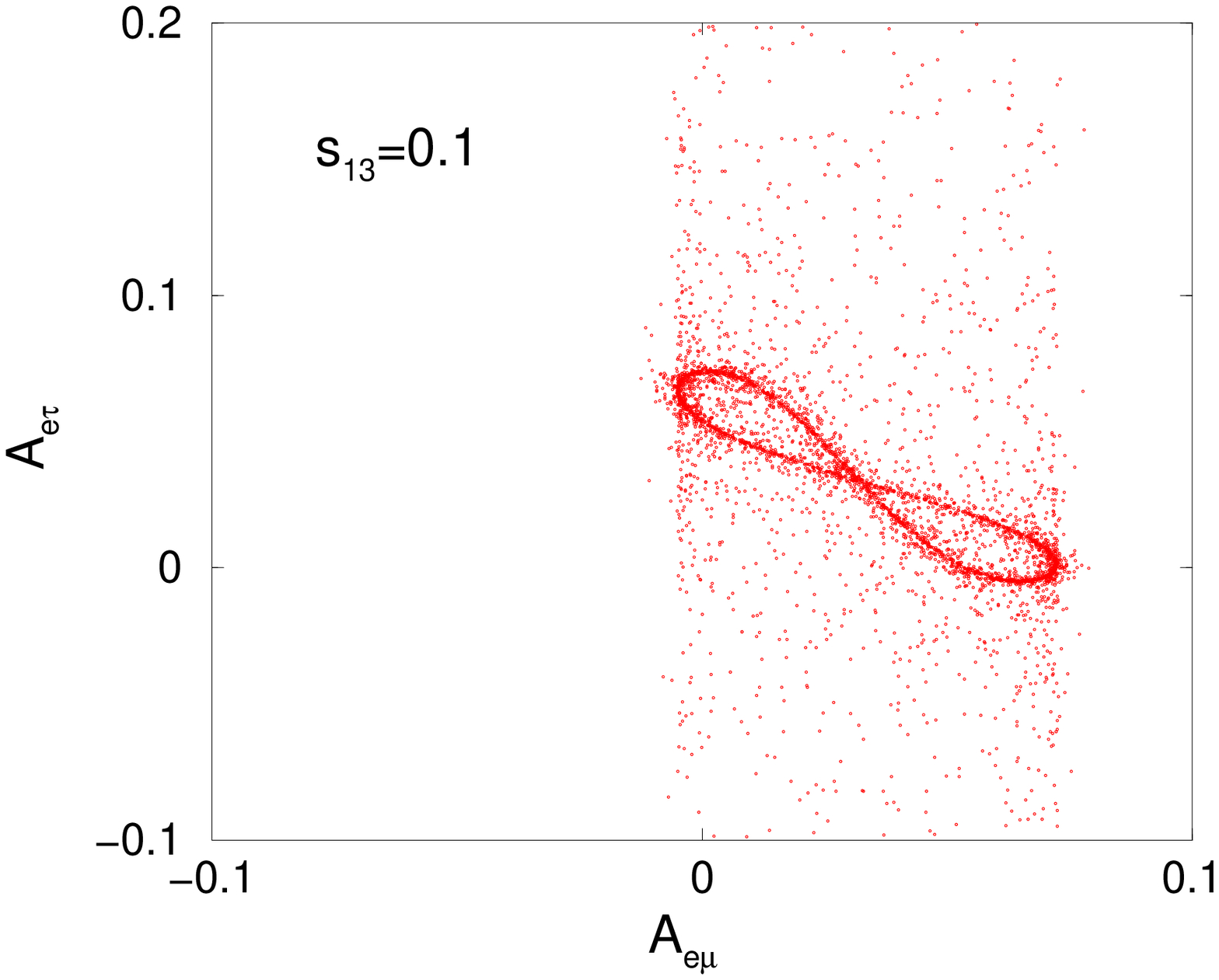} \includegraphics[width=0.49\linewidth]{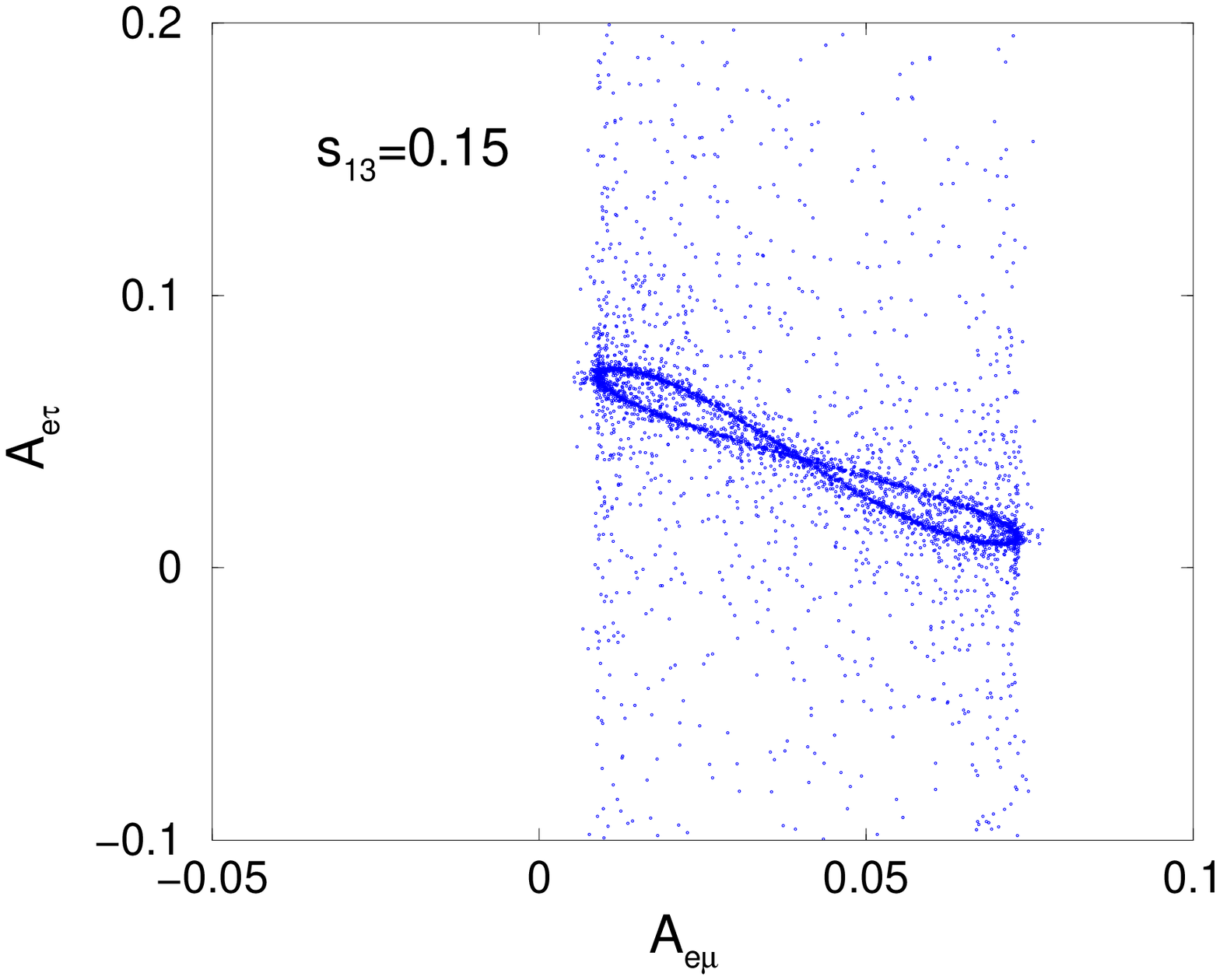}
\caption{\label{fig:aemaet} \it The asymmetry $A_{e\tau}$ plotted as a function of $A_{e\mu}$ for 
$\sin{\theta_{13}}= 0.02,~0.05,~0.10,~0.15$. We have assumed that the sign of $\Delta_{31}$ is positive (normal hierarchy), that $\theta_{23}$ is maximal and $\sin^2{\theta_{12}}=1/3$.
We have also fixed the mass differences to $\Delta m^2_{21}=8\times 10^{-5}$ eV$^2$   and $\Delta m^2_{31}=2.4\times 10^{-3}$ eV$^2$.
 In each panel we use $L=1500$ Km and $E_\nu=30$ GeV.
Notice the different x,y-axes scale in each plot. In each panel the marked closed line describes the standard model relation between the two asymmetries, while the dots span the domain obtained by varying the MUV parameters.}
\end{figure}

\begin{figure}[h!]
\centering 
\includegraphics[width=0.49\linewidth]{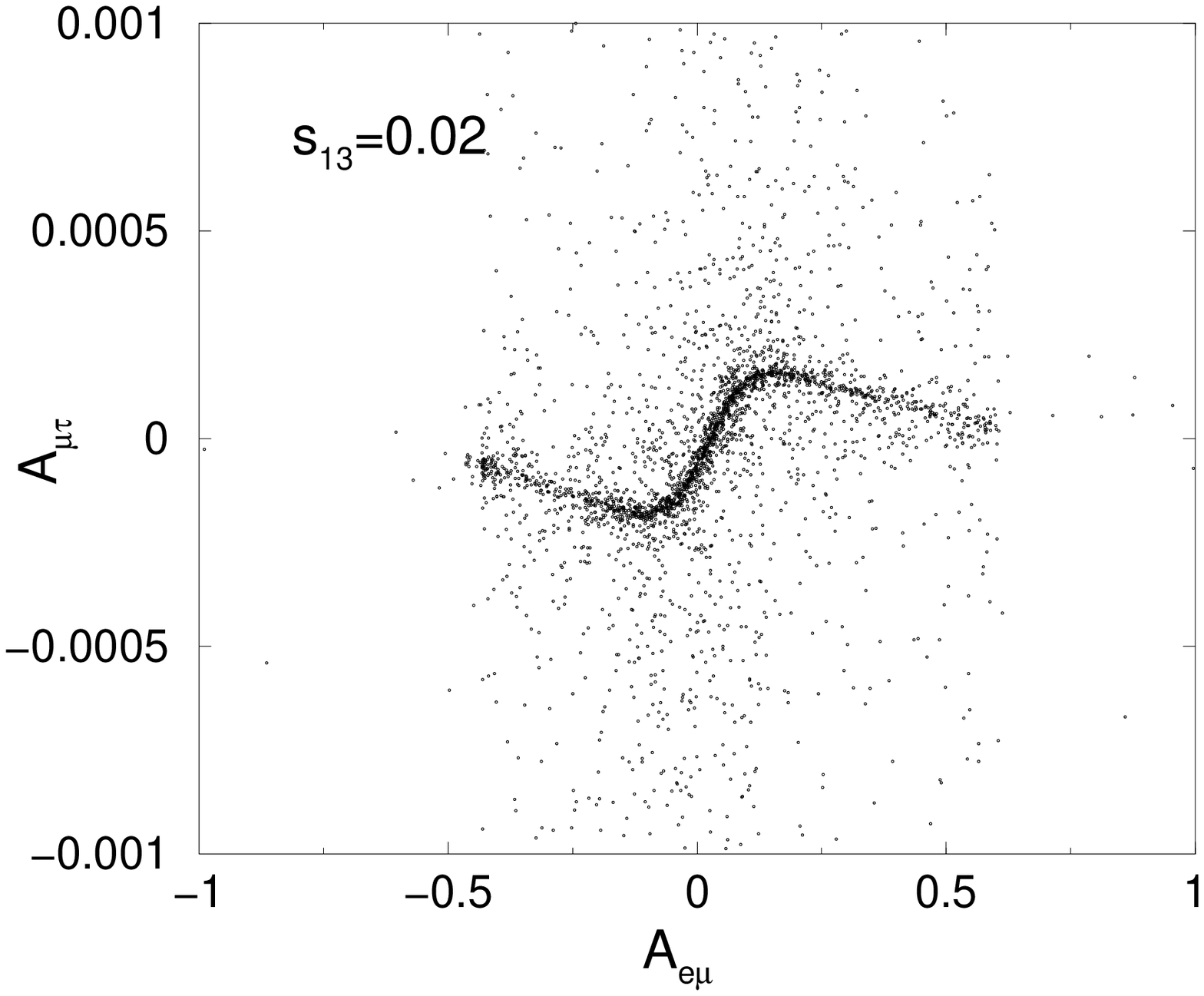} \includegraphics[width=0.49\linewidth]{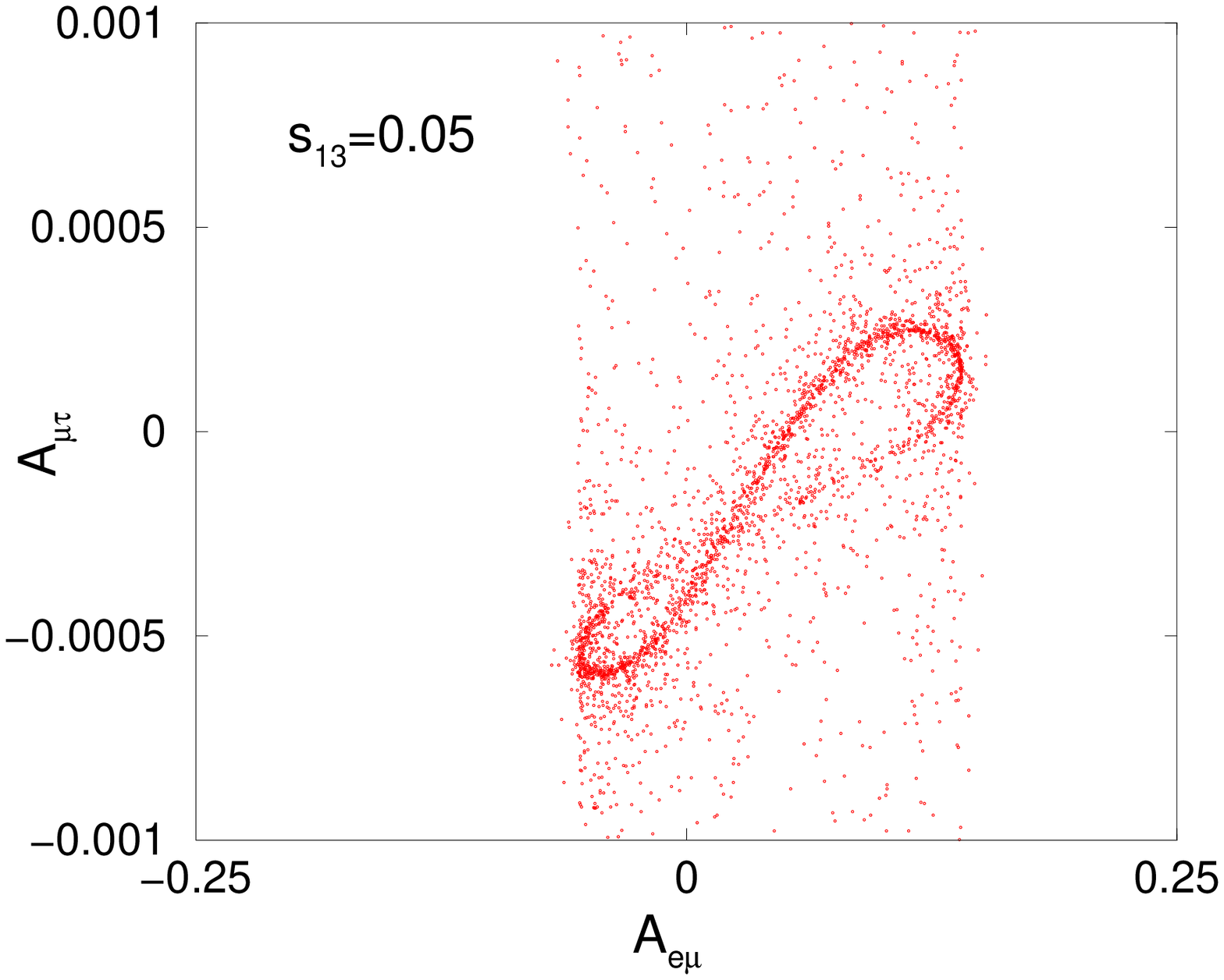}\\
\includegraphics[width=0.49\linewidth]{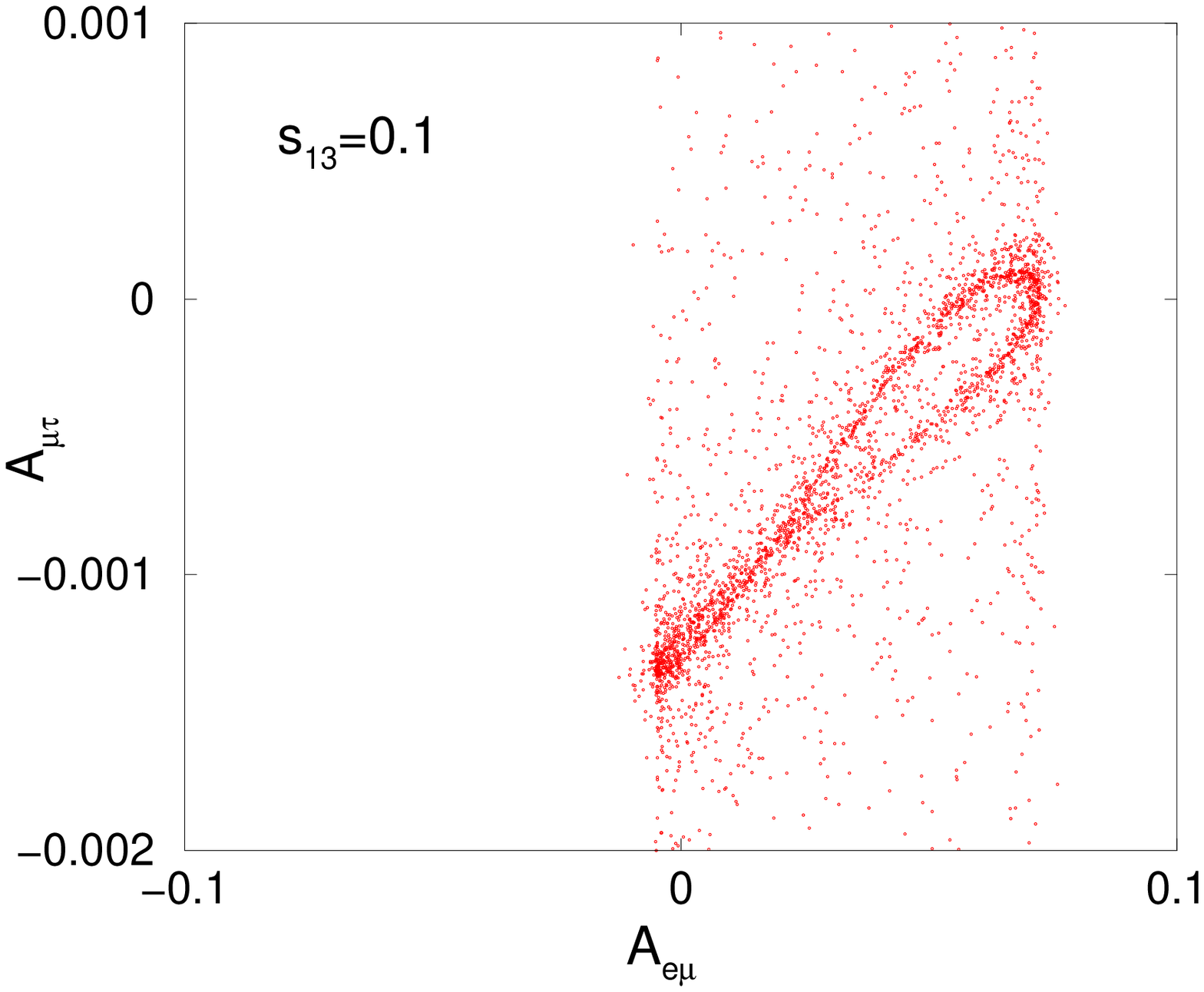} \includegraphics[width=0.49\linewidth]{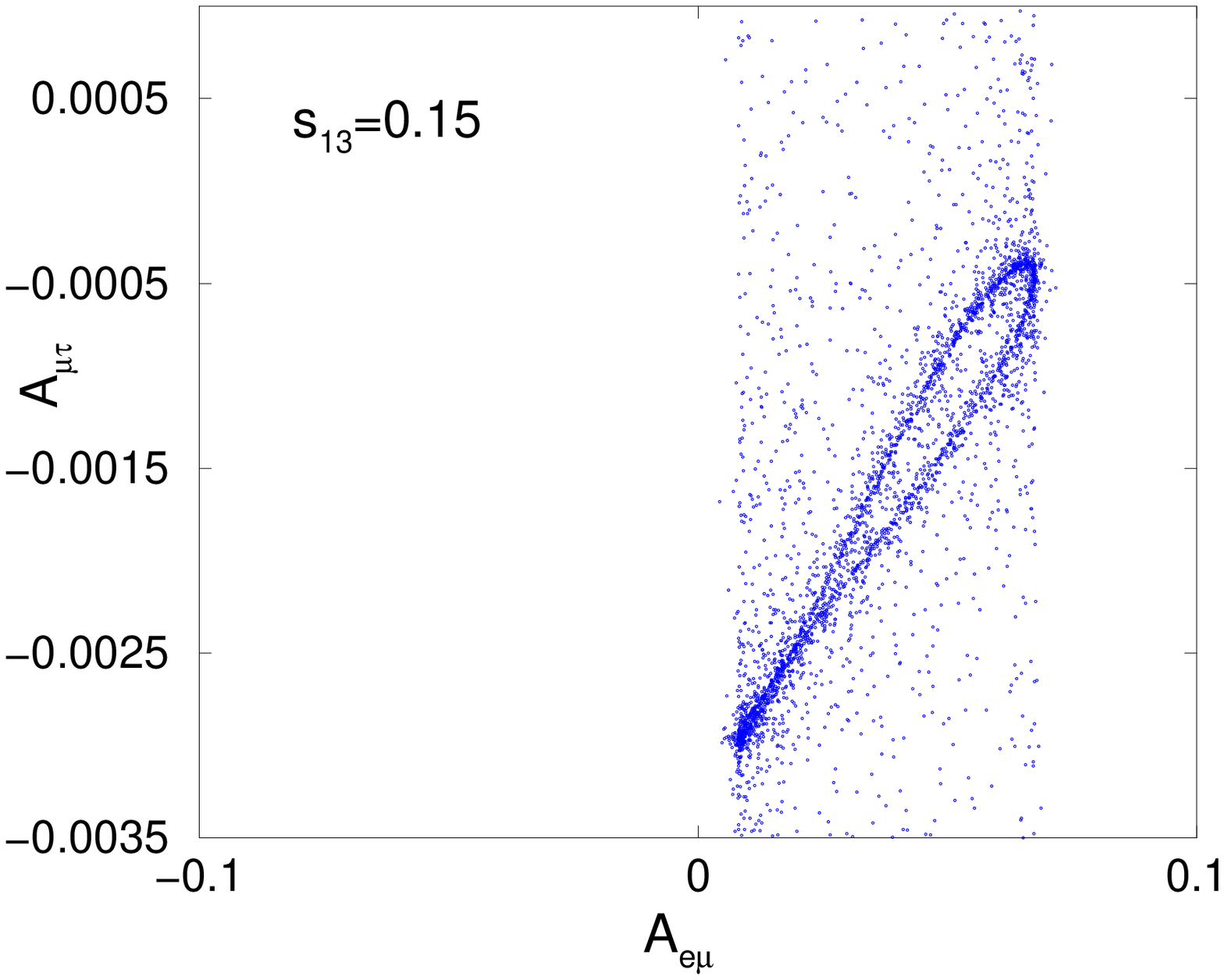}
\caption{\label{fig:aemamt} \it The asymmetry $A_{\mu\tau}$ plotted as a function of $A_{e\mu}$ for 
$\sin{\theta_{13}}= 0.02,~0.05,~0.10,~0.15$. The values of the other parameters are as in the previous figures. We compute   $A_{e\mu}$ for $E_\nu=30$ GeV  and
 $A_{\mu\tau}$ for $E_\nu=35$ GeV, both at $L=1500$ Km.
Notice the different x,y-axes scale in each plot. In each panel the marked closed line describes the standard model relation between the two asymmetries, while the dots span the domain obtained by varying the MUV parameters.  }
\end{figure}

In Figs.\ref{fig:Aem}, \ref{fig:Amt} and \ref{fig:Aet} we show the domain
spanned when the MUV parameters are moved in their allowed ranges, for each of
the
three asymmetries (in absolute value) and $\sin{\theta_{13}}=0.1$.
The asymmetry $A_{e\mu}$ can be  measured at both $\beta$-beams and Neutrino Factories whereas the Super-Beams are better suited to measure $A_{\mu e}$; in particular, we illustrate the effect of considering the MUV
parameters at NF@1500, T2HK, HE$\beta$B  and SPL@732, which have quite different
values of $L/\langle E_\nu \rangle$ (from $L/\langle E_\nu \rangle \sim 50$
Km/GeV for NF@1500 up to $L/\langle E_\nu \rangle \sim 2440$ Km/GeV for
SPL@732; notice also that T2HK, SPL@130 and LE$\beta$B share almost the same $L/\langle E_\nu \rangle \sim \mathcal O(300-400)$).
We can easily see that the prediction of $A_{e\mu}$ at each $\delta$ value is not much affected by varying the MUV parameters in their experimental allowed range at any of the facilities that we considered. 

\begin{figure}[h!]
\centering 
\includegraphics[width=0.8\linewidth]{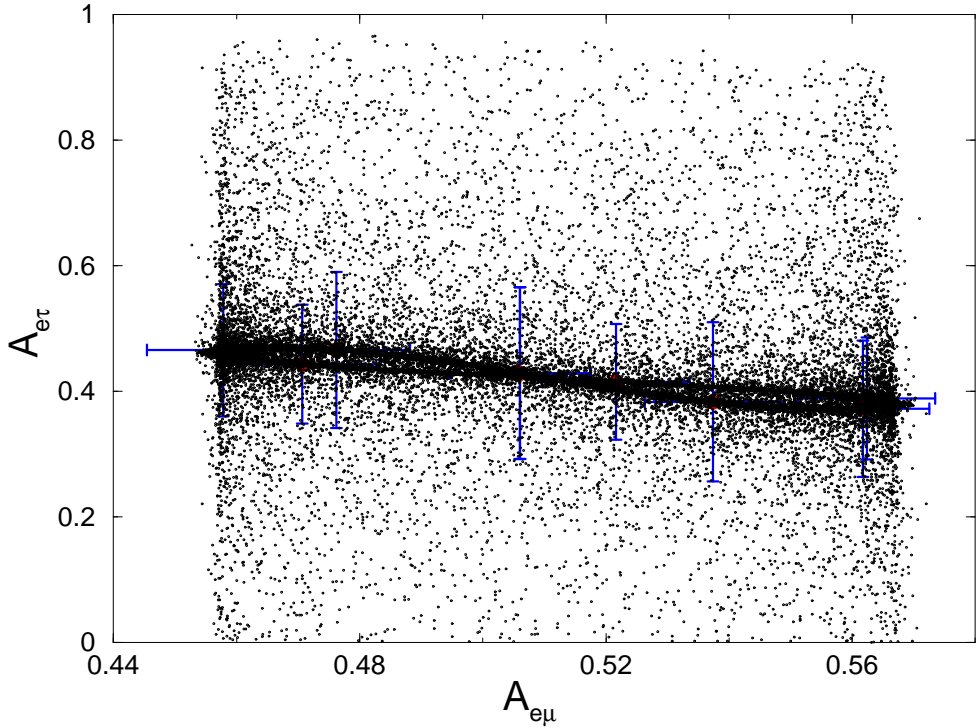}
\caption{\label{fig:aemaetRATE} \it The asymmetry $A_{e\tau}$ plotted as a function of $A_{e\mu}$ for 
$\sin{\theta_{13}}=0.10$, for a Neutrino Factory with $L=1500$ Km. The values of the other parameters are as in the previous figures. The solid line refers to the standard model prediction, with the uncertainties for both the asymmetries evaluated at some representative values of the phase $\delta$. Dots represent the spread of MUV predictions for the same quantities.
}
\end{figure}

This is not the case  for the asymmetries $A_{\mu\tau}$ and $A_{e\tau}$: while in the standard picture they are bounded in absolute value, these bounds can be grossly  violated in the MUV model.
In particular, the measurement of $A_{\mu\tau}$ can in principle be done at Neutrino Factories with different $L/\langle E_\nu \rangle$;
as we can see in Fig.\ref{fig:Amt}, the strongest effects of new physics are
obtained at the facility with the smallest $L/\langle E_\nu \rangle$, thus
showing that the Neutrino Factories with small baselines and large neutrino energies are better suited to detect new physics effects in the  $\nu_\mu \to \nu_\tau$ channel \cite{gavela}.

\begin{figure}[h!]
\centering 
\includegraphics[width=0.8\linewidth]{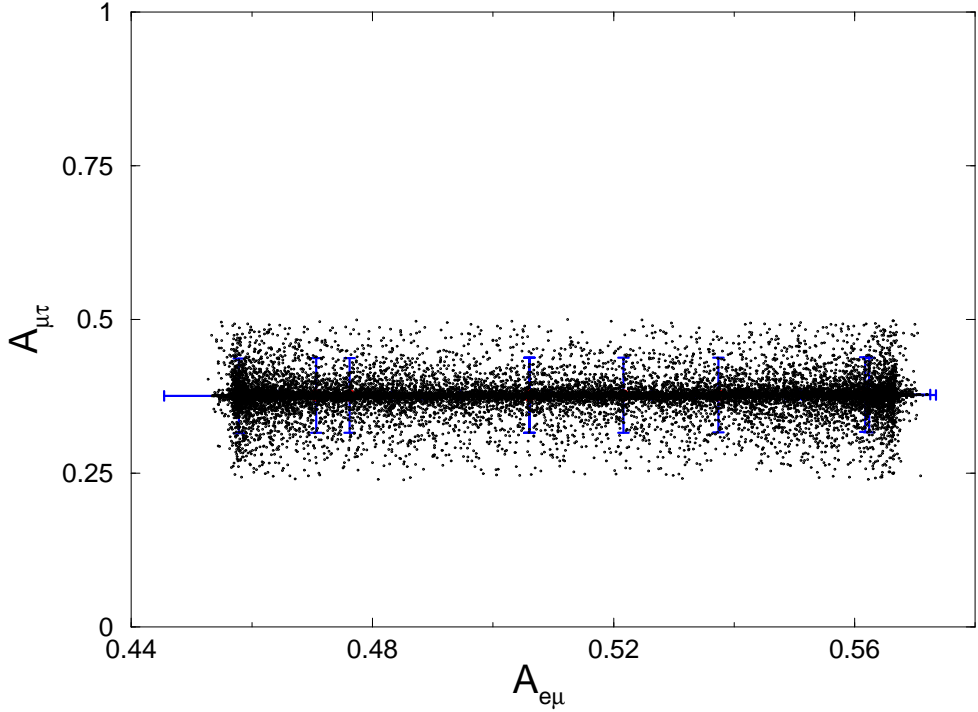}
\caption{\label{fig:aemamtRATE} \it The asymmetry $A_{\mu\tau}$ plotted as a function of $A_{e\mu}$ for
$\sin{\theta_{13}}=0.10$, for a Neutrino Factory with $L=1500$ Km. The values of the other parameters are as in the previous figures. The solid line refers to the standard model prediction, with the uncertainties for both the asymmetries evaluated at some representative values of the phase $\delta$. Dots represent the spread of MUV predictions for the same quantities.
}
\end{figure}

Finally, in Fig.\ref{fig:Aet} we present the results for the asymmetry
$A_{e\tau}$.
A part from the matter effects which modify the shape of the asymmetry, there is
not much difference between the two scenarios,
as a consequence of the fact that  the values of $L/\langle E_\nu \rangle$
differ at much by a factor of 3.

\begin{figure}[h!]
\centering 
\includegraphics[width=0.49\linewidth]{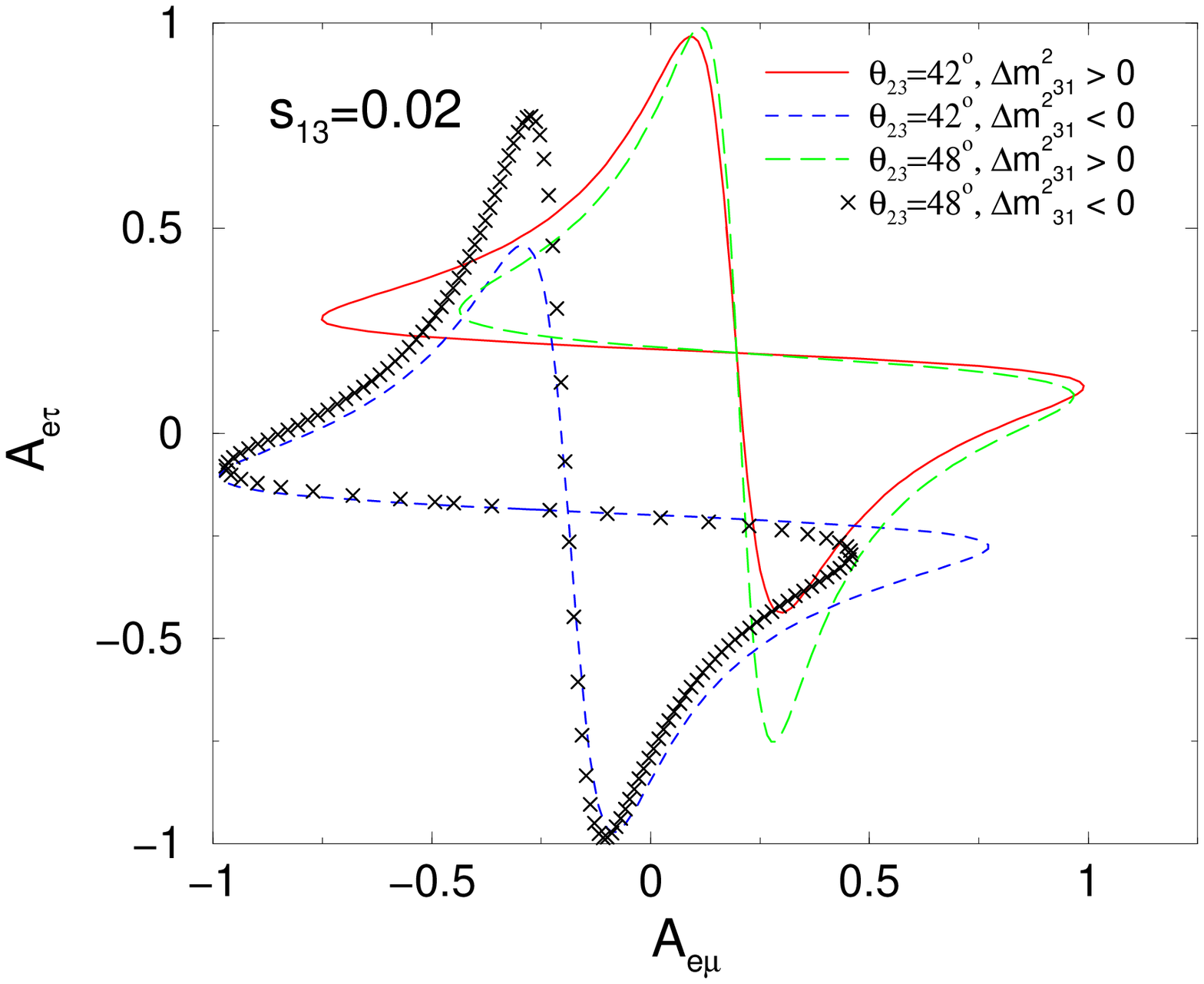}
\includegraphics[width=0.49\linewidth]{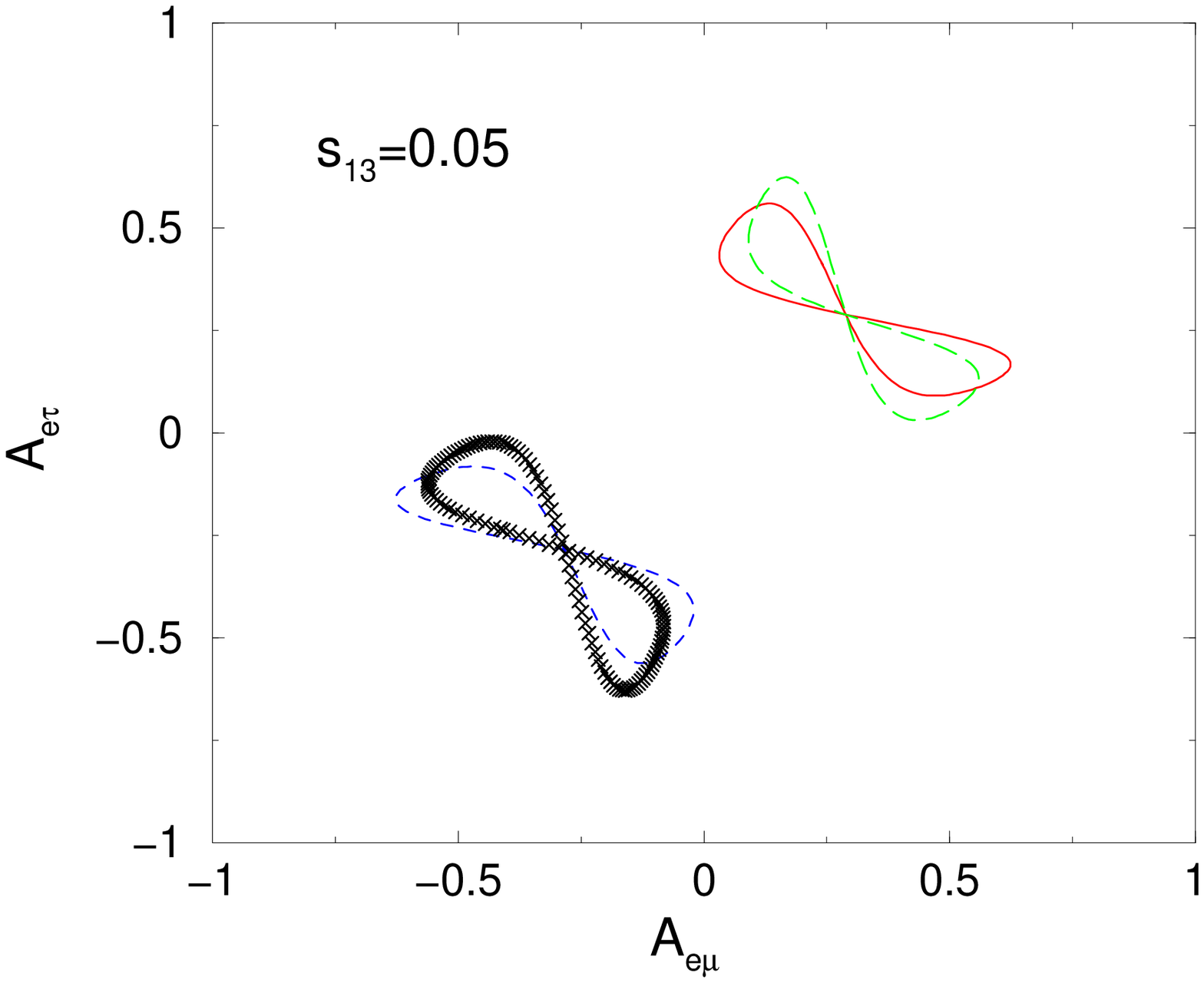}\\
\includegraphics[width=0.49\linewidth]{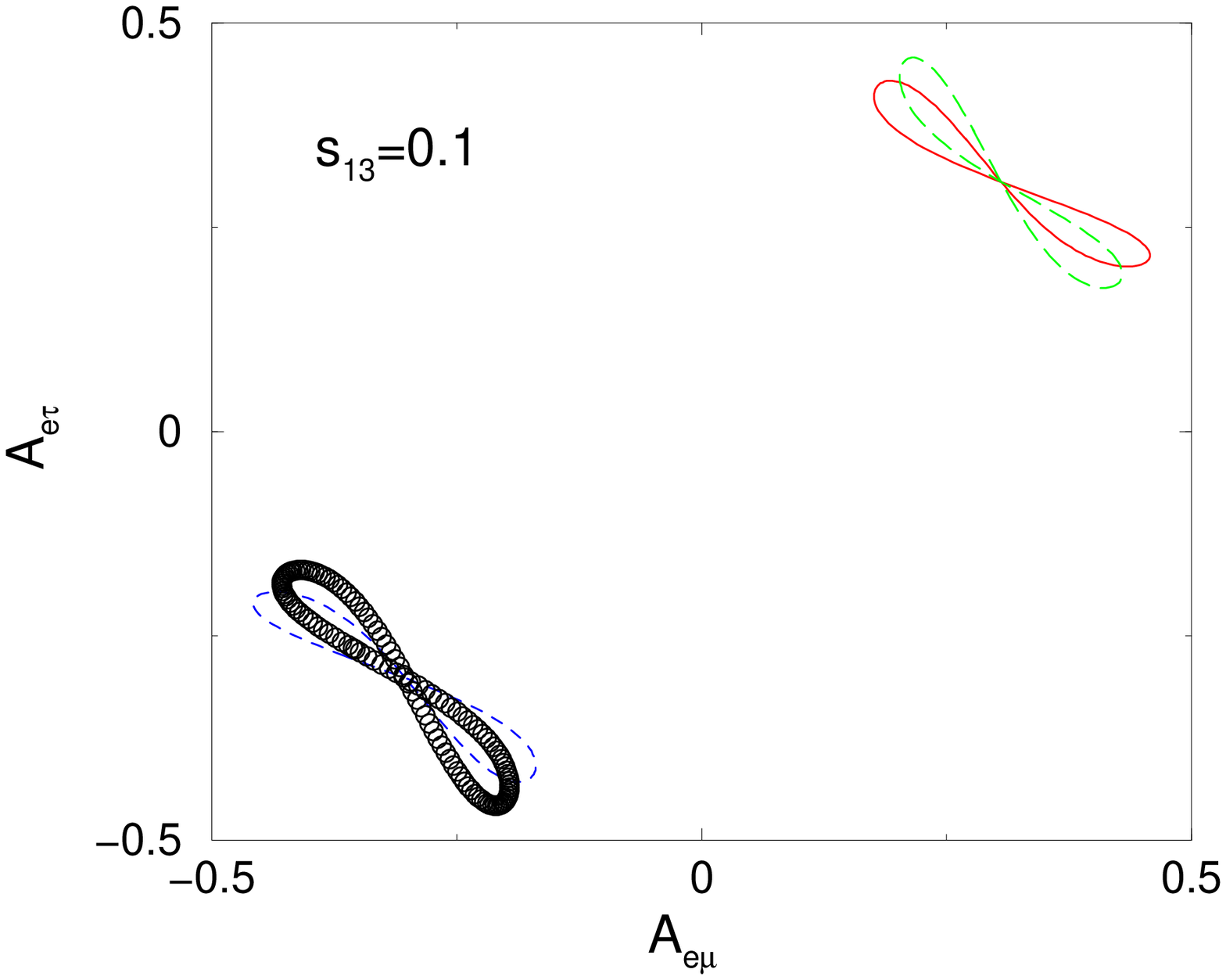}
\includegraphics[width=0.49\linewidth]{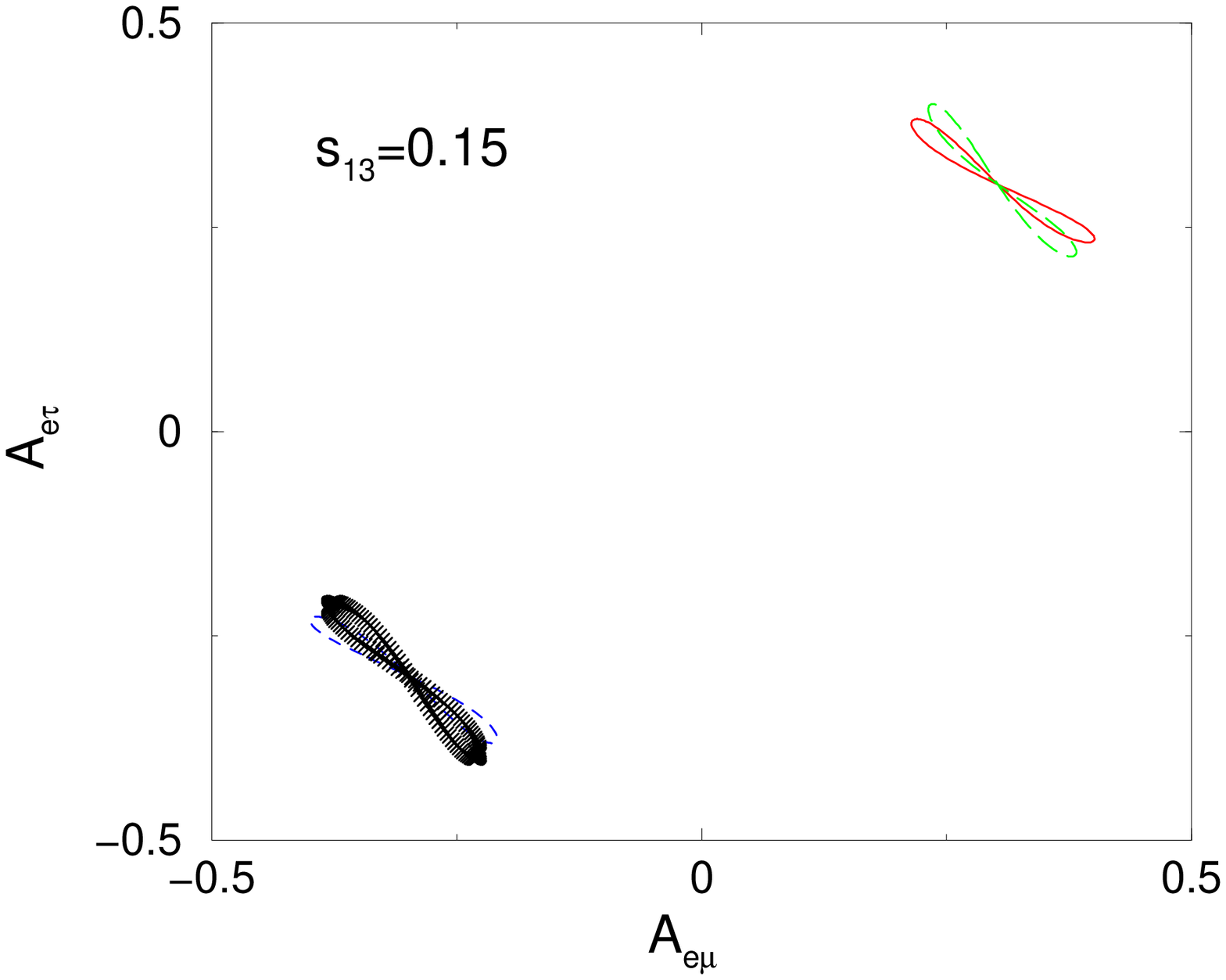}
\caption{\label{fig:deg} \it Effect of the octant and sign degeneracies on the
$A_{e\mu}$ and $A_{\mu\tau}$ asymmetries, for
$\sin{\theta_{13}}= 0.02,~0.05,~0.10,~0.15$. 
In each panel we used $L=4000$ Km and $E_\nu=30$ GeV. The solid line refers to $\theta_{23}=42^o$ and normal hierarchy, the dashed one to
$\theta_{23}=42^o$ and inverted hierarchy, the long-dashed to  $\theta_{23}=48^o$ and normal hierarchy and crosses refer to $\theta_{23}=48^o$ and inverted hierarchy.}
\end{figure}

\begin{figure}[h!]
\centering 
\includegraphics[width=0.6\linewidth]{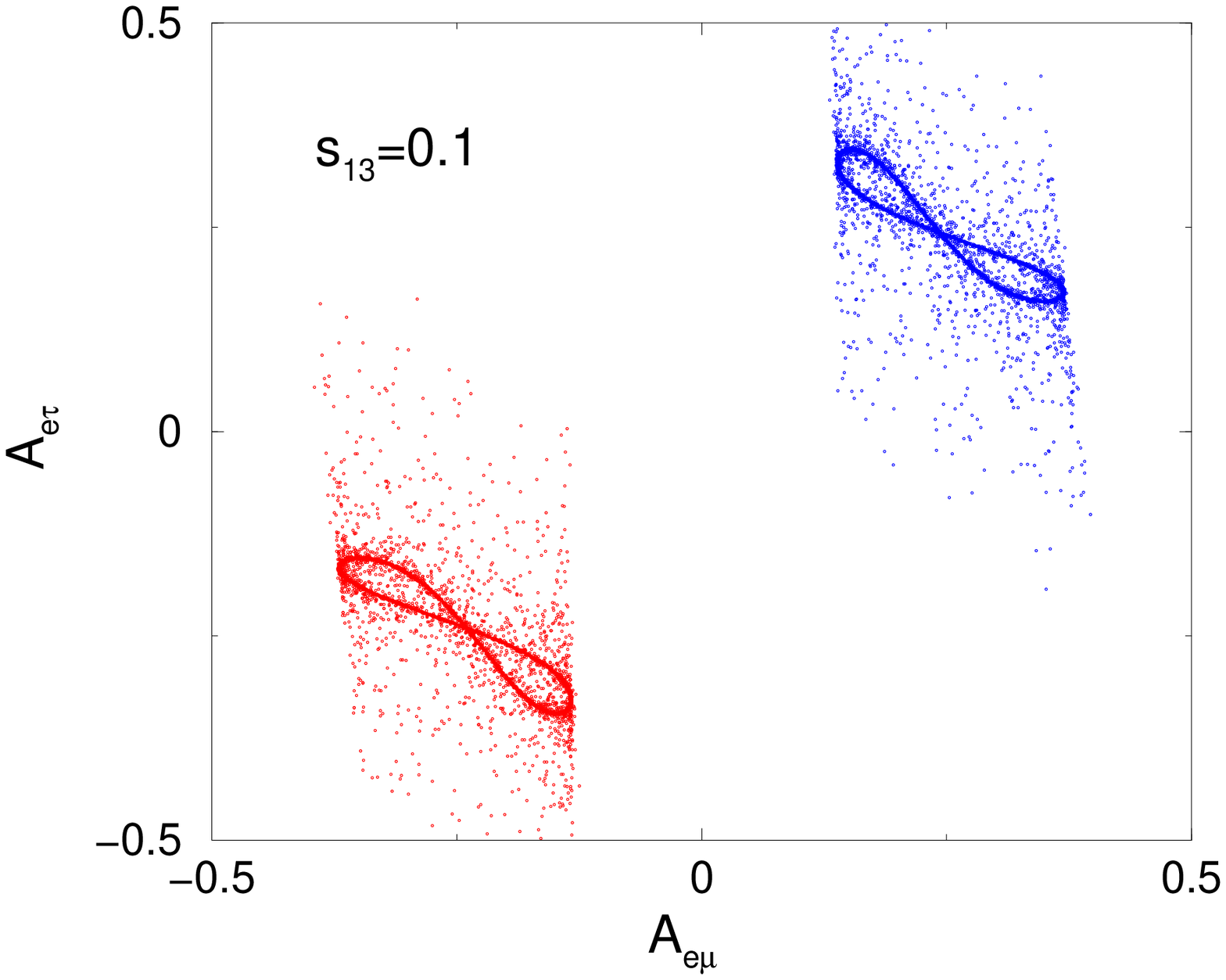}
\caption{\label{fig:degMUV} \it New physics effects in the MUV model in presence of the sign
degeneracy for $A_{e\mu}$ and $A_{e\tau}$ asymmetries, computed for
$\sin{\theta_{13}}=~0.1$ at $L=4000$ Km and $E_\nu=30$ GeV. Solid lines refer to the standard model scenario whereas dots are the MUV predictions. }
\end{figure} 

Clearly if the experimental values of one or more asymmetries are not compatible with the standard bounds, the PMNS model is directly disproved. Instead  we consider here the case where the measured values are compatible with the bounds. From the measured value of  $\delta$ the prediction for all CP violating asymmetries is obtained  in the standard model. It is then interesting to eliminate $\delta$ and plot each asymmetry as a function of the measured value of $A_{e\mu}$, which is the least affected by the MUV new physics. The resulting plot is in the form of a closed line. If the standard model is valid the measured values for each pair of asymmetries should fall on the line.  These plots are shown in the four panels in Fig.\ref{fig:aemaet} for $A_{e\tau}$ vs $A_{e\mu}$ and in Fig.\ref{fig:aemamt}  for $A_{\mu \tau}$ vs $A_{e\mu}$ at $L=1500$ Km and $E_\nu=30$ GeV (for $A_{e\mu}$ and
$A_{e\tau}$) or $E_\nu=35$ GeV (for $A_{\mu\tau}$), always for $\sin{\theta_{13}}= 0.02,~0.05,~0.10,~0.15$, a positive sign of $\Delta_{31}$, $\theta_{23}$  maximal and $\sin^2{\theta_{12}}=1/3$. These plots show that even with moderate precision a meaningful test of the PMNS mechanism of CP violation can be achieved. In fact the spread which is still allowed by the existing bounds on the MUV parameters is quite large for $A_{e\tau}$ and $A_{\mu \tau}$. In more general frameworks for new physics the deviations can even be larger. In particular the interval of allowed values for $A_{e\mu}$ could be wider.

\begin{figure}[h!]
\centering 
\includegraphics[width=0.49\linewidth]{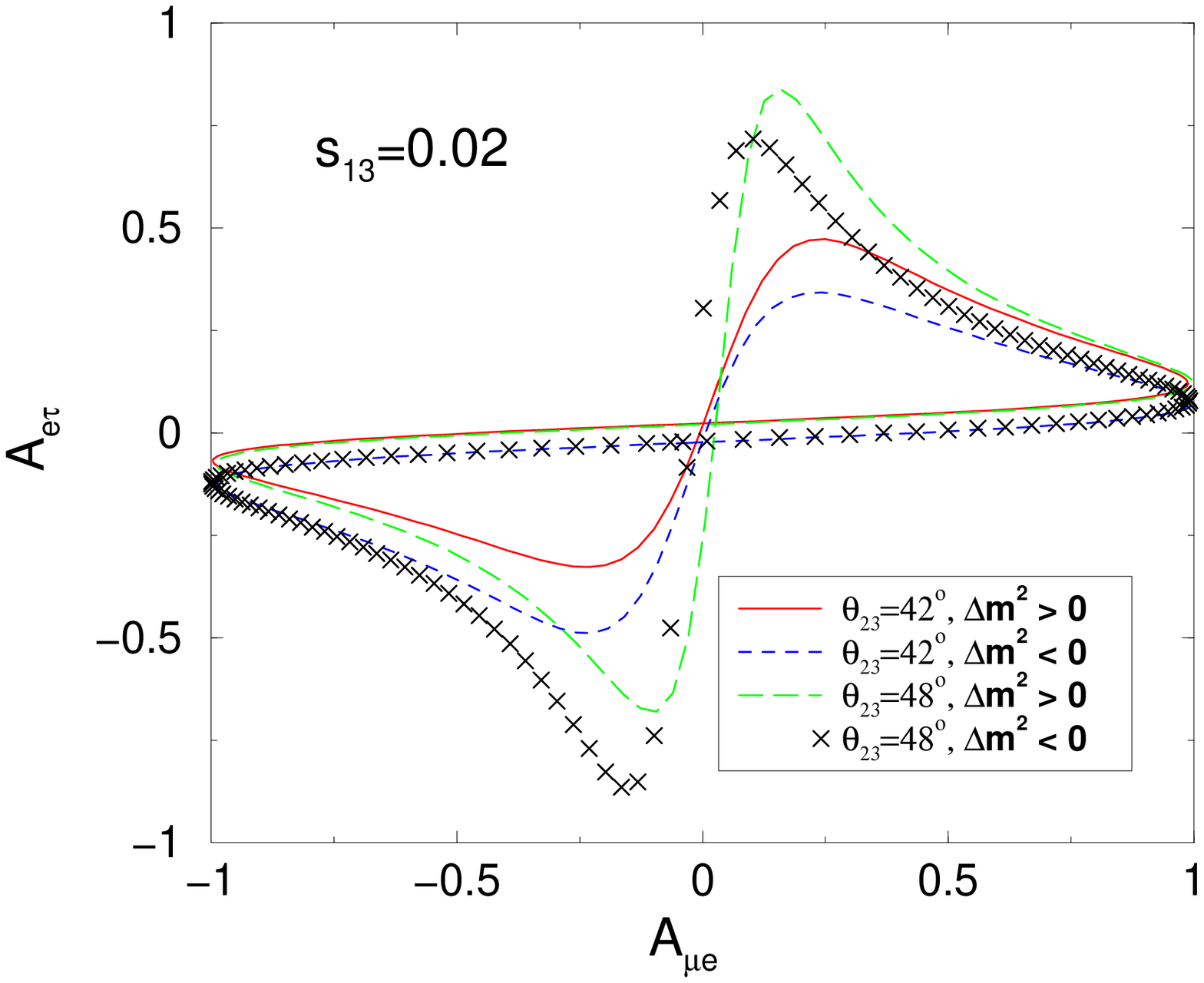}
\includegraphics[width=0.49\linewidth]{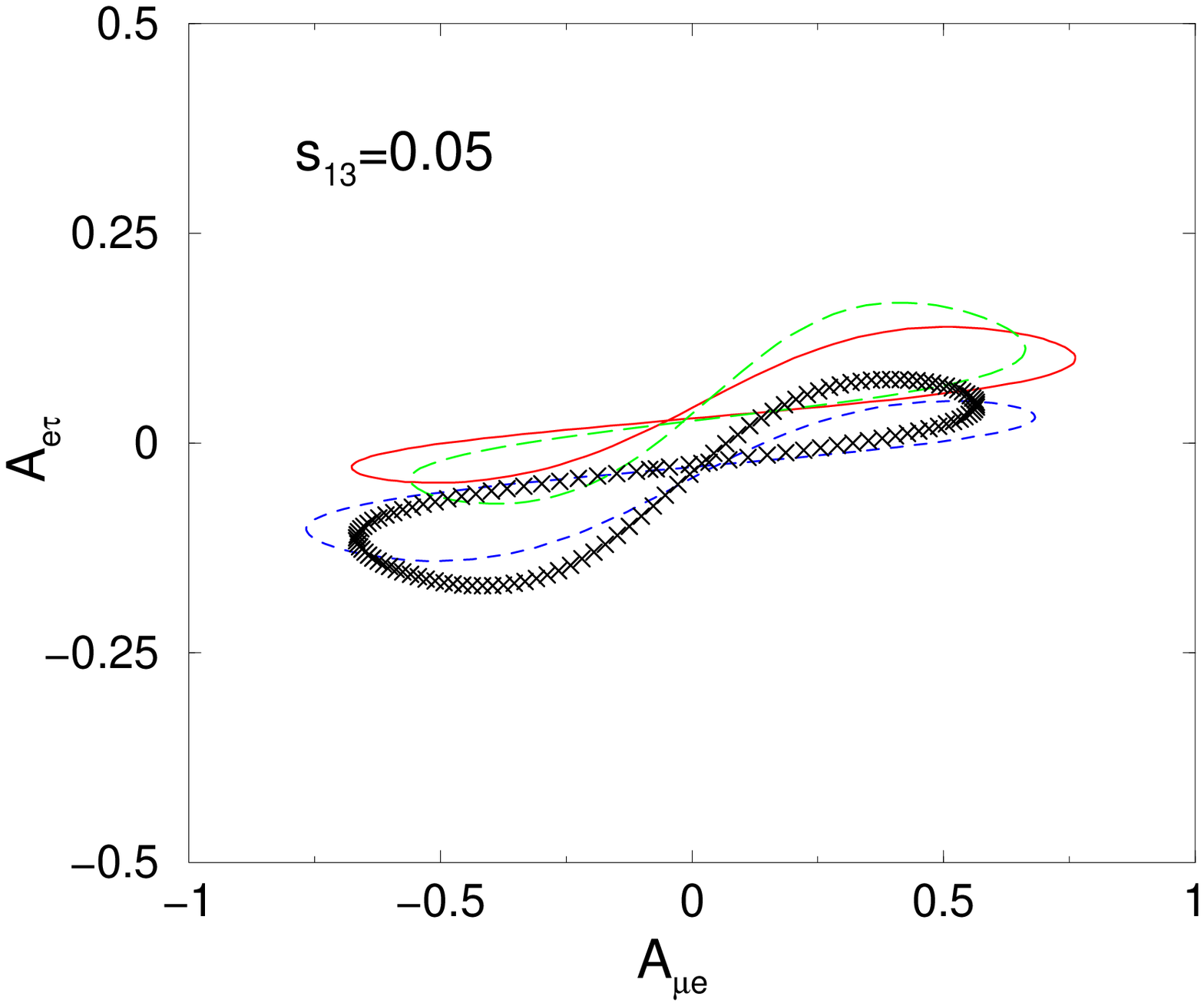}\\
\includegraphics[width=0.49\linewidth]{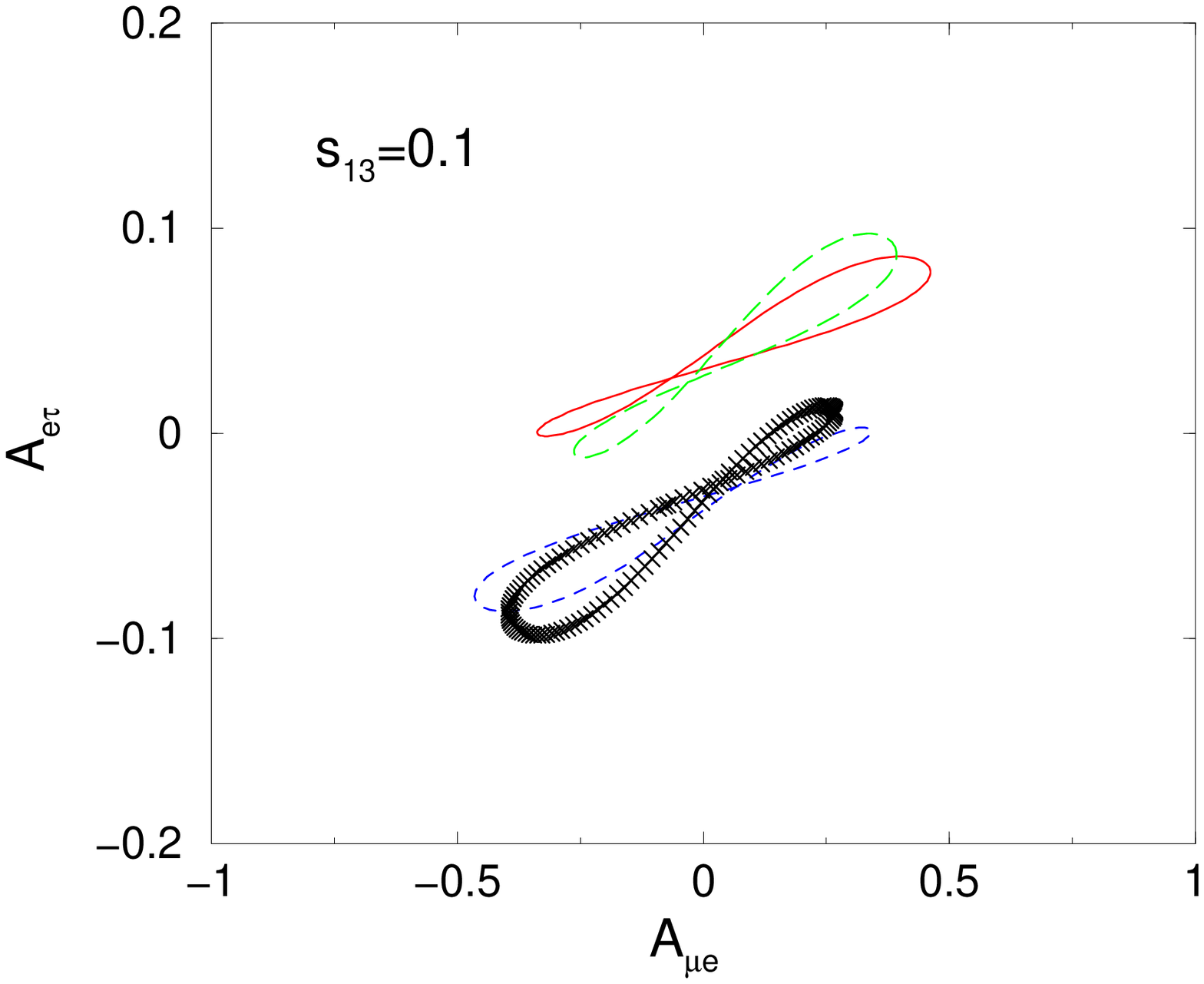}
\includegraphics[width=0.49\linewidth]{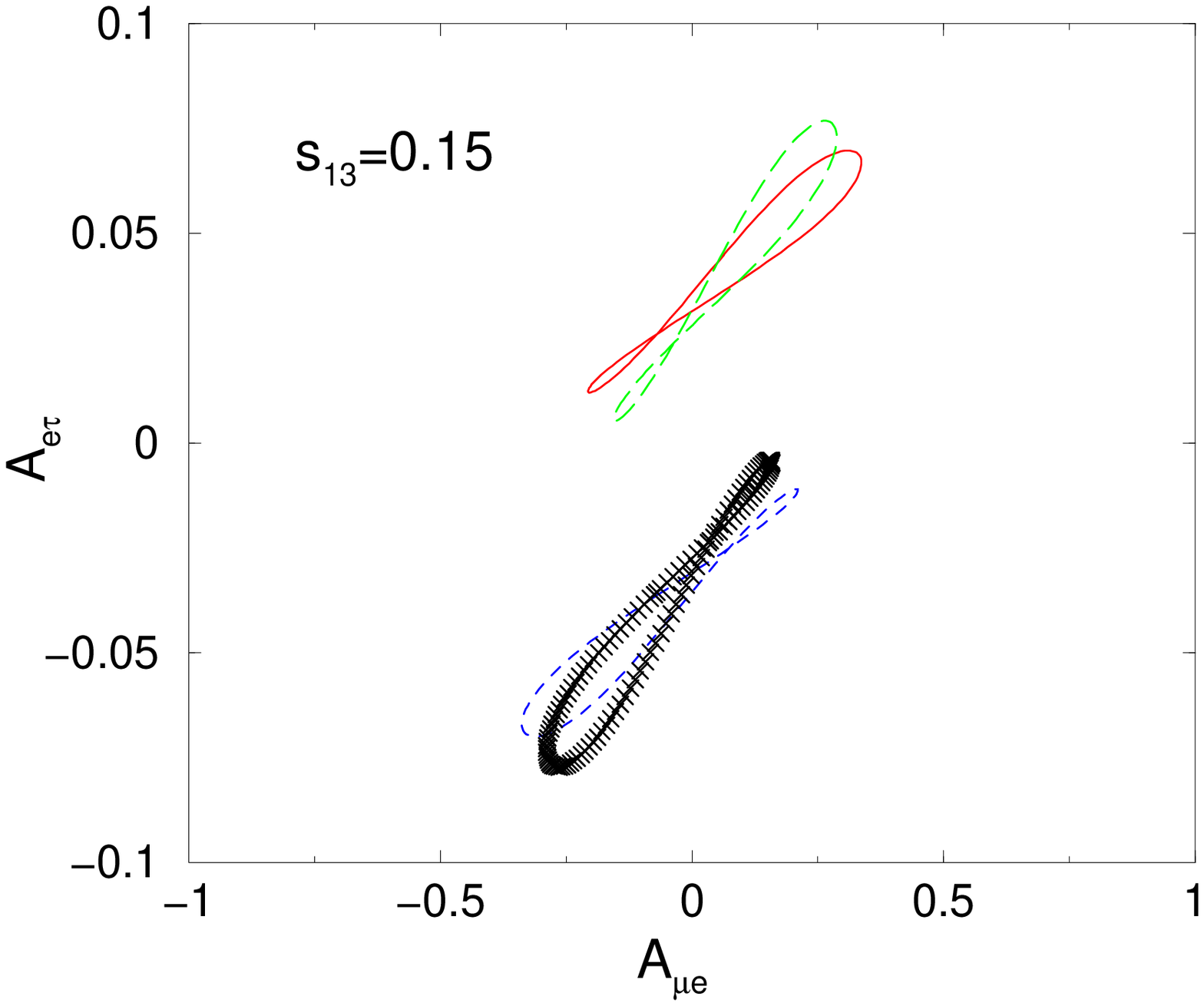}
\caption{\label{fig:deg2} \it Effect of the octant and sign degeneracies on the
$A_{\mu e}$ and $A_{\mu\tau}$ asymmetries, for
$\sin{\theta_{13}}= 0.02,~0.05,~0.10,~0.15$. 
$A_{\mu e}$ is computed for $L=295$ Km and $E_\nu=0.75$ GeV and $A_{e\tau}$ is computed for $L=1500$ Km and $E_\nu=30$ GeV. The solid line refers to $\theta_{23}=42^o$ and normal hierarchy, the dashed one to
$\theta_{23}=42^o$ and inverted hierarchy, the long-dashed to  $\theta_{23}=48^o$ and normal hierarchy and crosses refer to $\theta_{23}=48^o$ and inverted hierarchy.}
\end{figure}

\begin{figure}[h!]
\centering 
\includegraphics[width=0.6\linewidth]{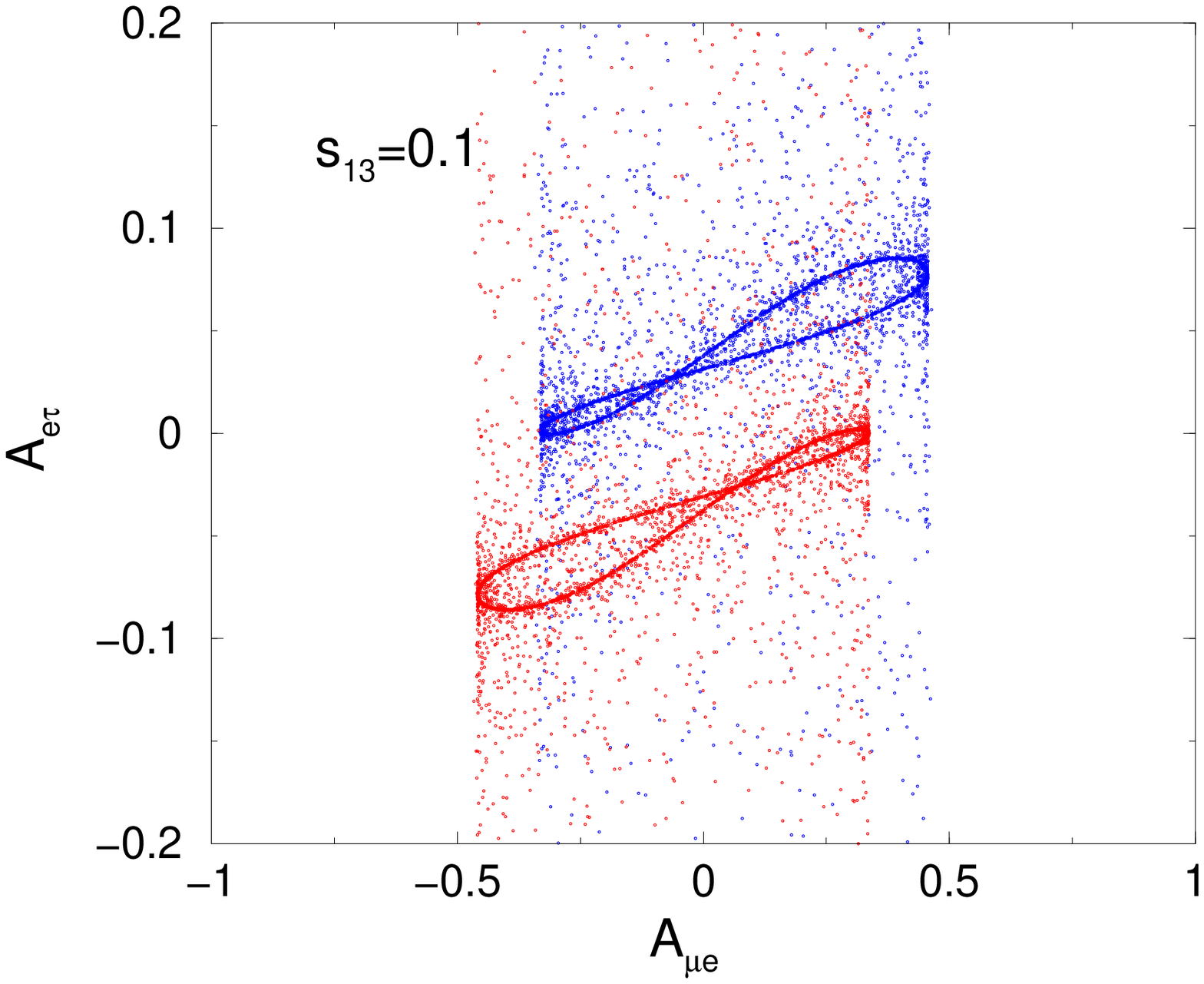}
\caption{\label{fig:degMUV2} \it
New physics effects in the MUV model in presence of the sign
degeneracy for $A_{\mu e}$, computed for $L=295$ Km and $E_\nu=0.75$ GeV, and $A_{e\tau}$ asymmetries, computed for
$L=4000$ Km and $E_\nu=30$ GeV, at fixed $\sin{\theta_{13}}=~0.1$. Solid lines refer to the standard model scenario whereas dots are the MUV predictions.}
\end{figure}

\subsection{A discussion of the expected experimental accuracy}

A central question now is related to the experimental capability to get a good enough  measurement of the asymmetries at the facilities that we consider: if the CP violating quantities will not be  measured  with sufficient precision, then the test can be invalidated because one cannot distinguish the deviation from the standard model results due to MUV effects.
To illustrate this point, we consider a neutrino factory with $L=1500$ Km and $E_\mu=50$ GeV and build the relevant {\it integrated } asymmetries from the number of expected events $N_\beta$ and $\bar N_\beta$ in a given detector according to:
\bea
\label{eq:asirate}
A_{\alpha \beta} &=&\frac{N_\beta-\bar N_\beta}{N_\beta+\bar N_\beta}
\eea 
where the event rates for the $\nu_\alpha \to \nu_\beta$ and the CP conjugate $\bar\nu_\alpha \to \bar\nu_\beta$ transitions are computed from:
\be
\label{eq:asirate2}
N_\beta =\int_{E_\nu} dE_\nu \,P_{\alpha \beta}(E_\nu)\,\sigma_\beta(E_\nu)\,\frac{d\phi_\alpha}{dE_\nu}(E_\nu) \,\varepsilon_\beta(E_\nu)
\ee
\be
\label{eq:asirate3}
\bar N_\beta =\int_{E_\nu} dE_\nu \,P_{\bar\alpha \bar\beta}(E_\nu)\,\sigma_{\bar\beta}(E_\nu)\,\frac{d\phi_{\bar\alpha}}{dE_\nu}(E_\nu) \,\varepsilon_\beta(E_\nu)
\ee
in which $\sigma_{\beta (\bar \beta)}$ is the cross section for producing the lepton $\beta(\bar \beta)$,  
$\varepsilon_{\beta(\bar \beta)}$ the detector efficiency to reveal that lepton and  $\phi_{\alpha(\bar \alpha)}$ the initial neutrino flux at the source.  The asymmetries we are interested in are $A_{e\mu}$, $A_{e\tau}$ and $A_{\mu\tau}$; then 
$\beta$ can be a muon or a tau (and their antiparticles).
We perform a simulation of a neutrino factory in the spirit of \cite{Cervera:2000kp}, taking
$1\cdot 10^{21}$ $\mu^+$ $\times$ 4 years, 
$1\cdot 10^{21}$ $\mu^-$ $\times$ 4 years \cite{ISS} and putting all the
detectors we need at the same baseline $L=1500$ Km.  It is obvious that different technologies are needed to detect muons and taus; for this reason,
we imagine to look for golden muons  using an improved version of the
MIND detector described in \cite{dydak} and, to evaluate the uncertainty associated with $A_{e\mu}$, we use the  efficiencies and backgrounds from \cite{anselmo2}. The detector mass is fixed to 50 Kton. 
The search of $\tau$ events takes place looking for its $\mu$ decay mode; in the following we use an Opera-like detector (the so-called Emulsion Cloud Chamber (ECC) detector \cite{opera}). With this detector, the {\it silver} channel $\nu_e \to \nu_\tau$ has been carefully studied in \cite{Donini:2002rm}, and we refer to these papers for our estimate of
efficiencies and backgrounds.  However, such a detailed study is still missing for the $\nu_\mu \to \nu_\tau$ transition in the context of the neutrino factory; here we follow \cite{gavela} and assume that it should also be possible to look for the other $\tau$ decay modes, thus gaining a factor 5 in statistics (at the prize of increasing also the backgrounds) \footnote{A similar option has been also investigated for the silver channel first in
\cite{Huber:2006wb} and further discussed in \cite{delellis}.}.
The mass of the detector is 10 Kton.
An important point to stress is that at the level of precision required to see new physics effects, systematic uncertainties play a relevant
role; in the following we assume a 
conservative overall 10\%  systematic error for the silver detector and an overall
2\% for the golden one. 
For the sake of illustration, in the following 
we integrate the rates in eqs.(\ref{eq:asirate2})-(\ref{eq:asirate3}) just over the whole $E_\nu$ range; this means that
the asymmetries do not profit of any spectral feature of the signal (certainly available with such good-performance detectors),  
which would help in extracting more accurate information on CP-violation. In this respect, our results can be improved in a more refined analysis which is beyond the scope of this section.
The results are shown in  Fig.\ref{fig:aemaetRATE}, where we plot the asymmetry $A_{e\tau}$ as  a function of $A_{e\mu}$ and their corresponding standard model uncertainties, at some representative value of the CP phase $\delta$, as well as the spread predicted in the MUV environment.  The value of $\theta_{13}$ is fixed to $\sin{\theta_{13}}= 0.10$. 
It can be clearly seen that  the asymmetry uncertainties are not as big as to spoil the possibility to see substantial deviations from the standard model predictions, since many of the values  obtained in the MUV framework are well beyond the asymmetry uncertainties
\footnote{Notice that the uncertainty on $A_{e\tau}$ is larger than that on $A_{e\mu}$, as a consequence of the
smaller event rates and larger background for the silver channel, although this is not clearly visible in the figure because of the  different horizontal and vertical scales used to make the plot more understandable.}.
For comparison, this scenario is equivalent to the third panel of Fig.\ref{fig:aemaet}; however, the numerical
values of the asymmetries are different, as a result of integrating the transition probabilities  over different neutrino and antineutrino cross sections which, at the relevant neutrino energies, are in the ratio $\sigma_{\beta}/\sigma_{\bar\beta}\sim \mathcal O$(2-3).

The same  can also be done with the other asymmetry, $A_{\mu\tau}$. This is illustrated in Fig.\ref{fig:aemamtRATE}, in which we plot $A_{\mu\tau}$ as  a function of $A_{e\mu}$ and their corrresponding uncertainties.   It is clear that this case looks less favourable, as while the size of the experimental errors is comparable the MUV scatter plot is more concentrated. Still, also in this plot, there are MUV predictions for $A_{\mu\tau}$ not falling inside the error band (whereas, as expected, the already small deviations for $A_{e\mu}$ are completely overshadowed by the uncertainty).  However, 
compared to Fig.\ref{fig:aemamt}, $A_{\mu\tau}$ is now substancially different from zero; in particular, the almost constant value is a direct consequence of the $\sigma_{\beta}$-to-$\sigma_{\bar\beta}$ ratio $\mathcal R$; in fact, assuming for simplicity that  $N_\beta\sim \mathcal R \,P$ and $\bar N_\beta\sim  \bar P$,   we can estimate $A_{\mu\tau}$ as follows:
\bea
\nn A_{\mu\tau} = \frac{(\mathcal R-1)P+(P-\bar P)}{\mathcal R P + \bar P} \sim \frac{(\mathcal R-1)}{(\mathcal R+1)}
\eea
where in the last step we have assumed that $P-\bar P \sim 0$ in the standard model.  Thus, $\mathcal R\sim \mathcal O$(2-3) justifies the almost constant value of $A_{\mu\tau}$  in Fig.\ref{fig:aemamtRATE}.
We stress that our error estimates can be improved in several ways; in particular, we have checked that the uncertainties on $A_{e\tau}$ and
 $A_{\mu\tau}$ are rather sensitive to the assumed systematic errors and this is especially important
if one want to amplify the MUV effects in the $\nu_\mu \to \nu_\tau$ transition.

As a general conclusion, this exercise shows that the indications that can be drawn by looking at the asymmetries at the probability level, computed at the peak energy of the neutrino spectra, are not invalidated when reliable error estimates are taken into account.  It is also true that, a part from the ECC detector, the proposed facilities have not been built yet and MonteCarlo simulations for the detector
response are still in progress. For these reasons, we prefer to investigate the other relevant issues of the paper using the {\it non-integrated} asymmetries, as done at the beginning of this section.

\subsection{The impact of degeneracies}

One may wonder what would be the impact of the sign and octant degeneracies on
the significance of the tests if the corresponding ambiguities were not already
fixed at the time of the CP violation experiment. This is studied in Figs.
\ref{fig:deg}-\ref{fig:degMUV}. In Fig.\ref{fig:deg} we see the possible impact
of the sign and the octant degeneracies for the asymmetries computed in the
standard picture for a very long baseline experiment with $L=4000~{\rm Km}$ (so
that the matter effect is particularly large and the sign degeneracy is
amplified) and $E_\nu=30~{\rm GeV}$. In this plot for the octant degeneracy we have
assumed that $\theta_{23}=42^o$. 
The results show that the ambiguity induced by an octant change is not very
important because the $\theta_{23}$ angle is in any case
close to maximal. The sign ambiguity instead makes a lot of difference at such a long
baseline distance. This implies that an experiment of this type can indeed
easily measure the $\Delta_{31}$ sign if this is not already known. In fact, as can be
seen from Fig.\ref{fig:degMUV} (where, for simplicity, the curves are plotted only for $\theta_{23}=42^0$) the new physics effects would not be large enough
to fill up the gap arising from the sign change, at least for not too small values of
 $\theta_{13}$. Conversely, the sign degeneracy does not spoil the possibility
of a meaningful test for the presence of new physics.

\begin{figure}[h!]
\centering 
\includegraphics[width=0.49\linewidth]{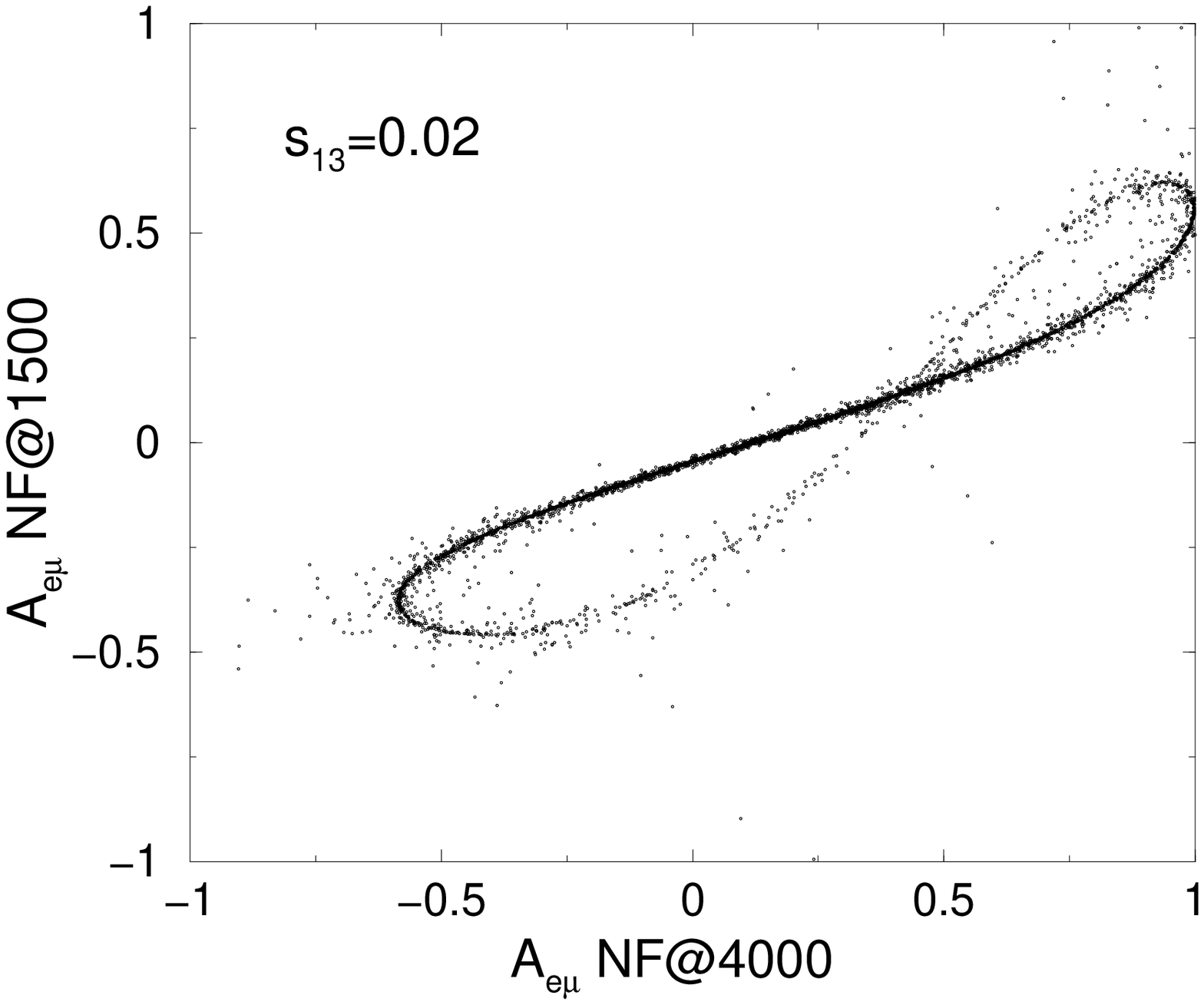}
\includegraphics[width=0.49\linewidth]{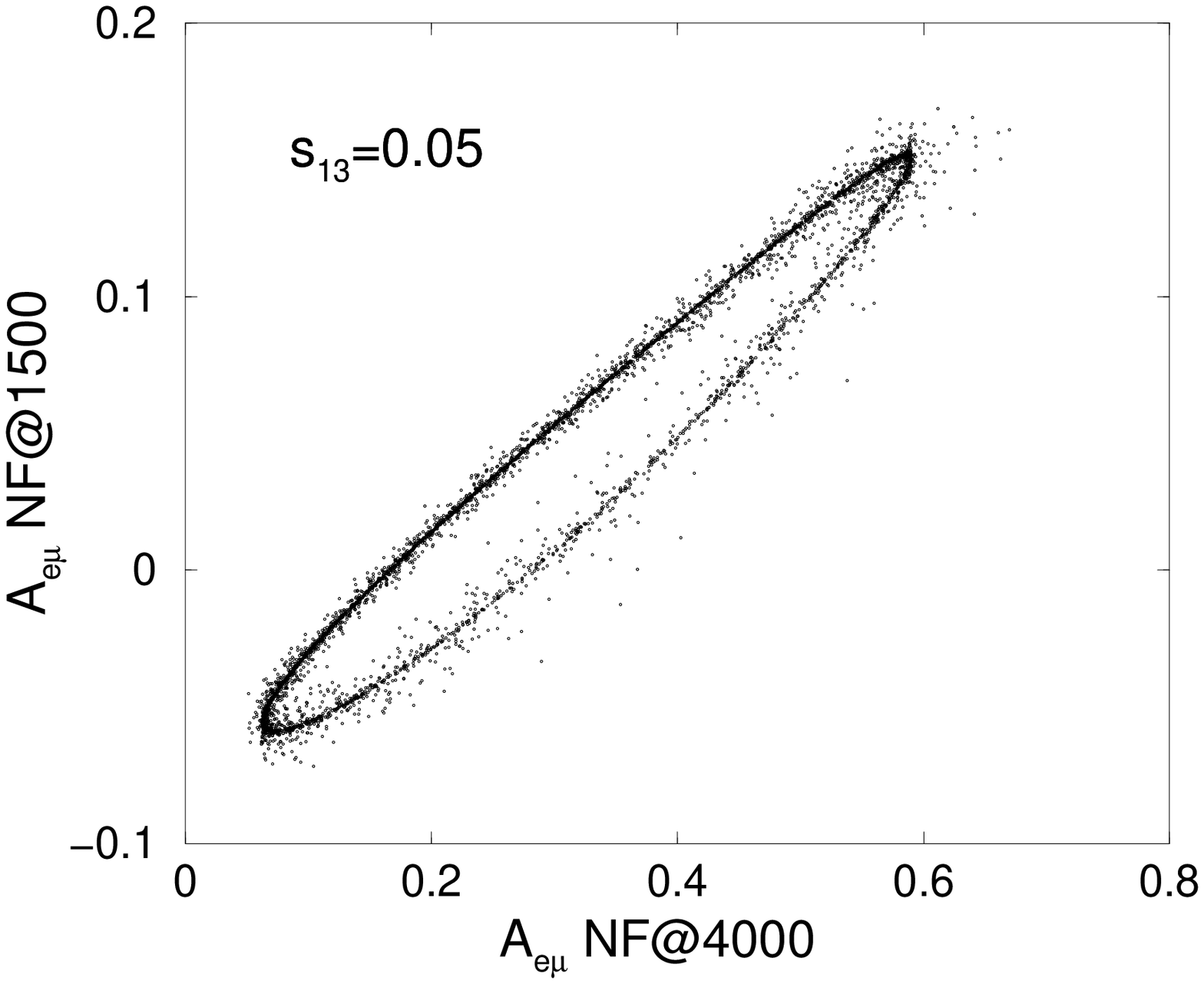}\\
\includegraphics[width=0.49\linewidth]{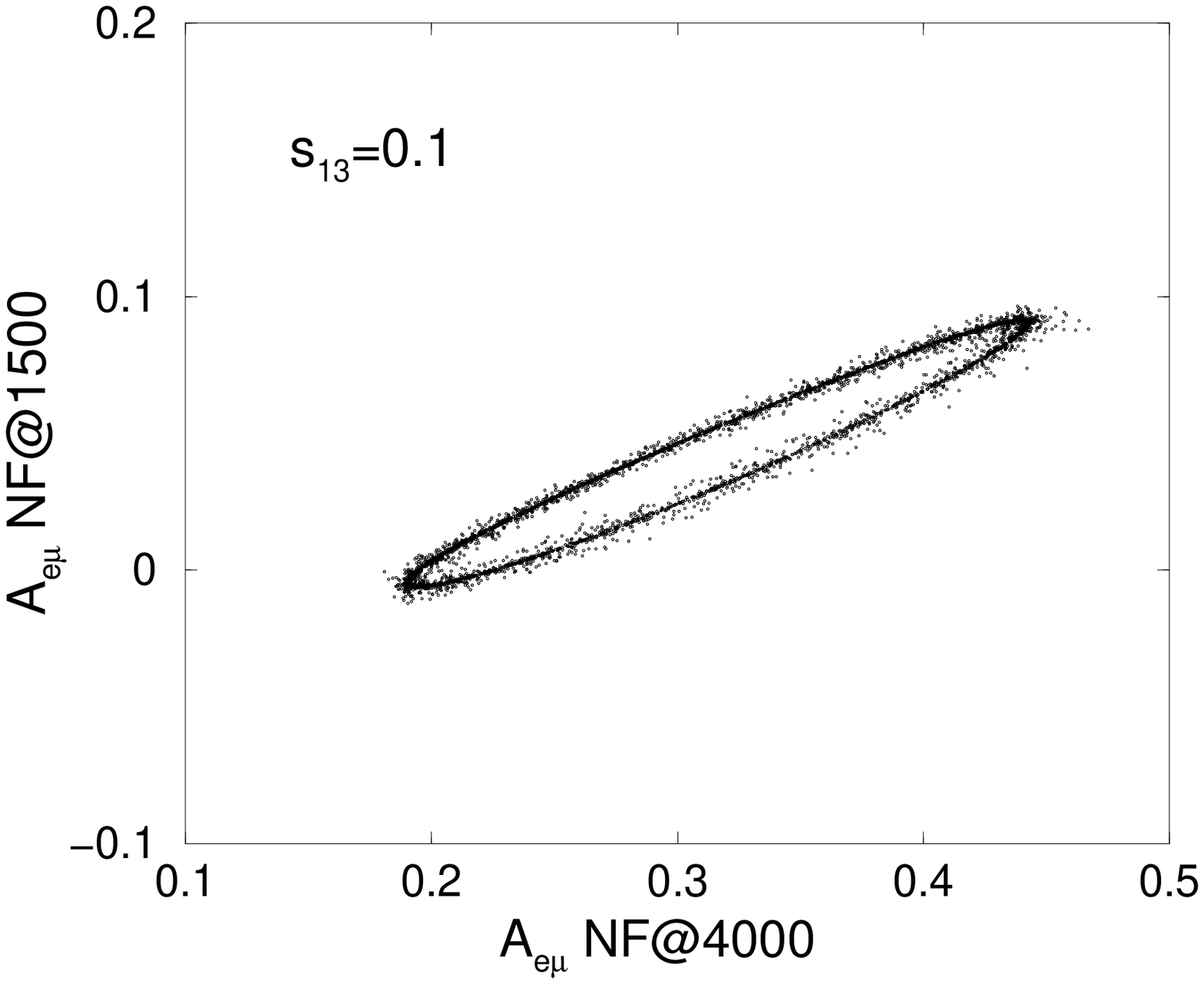}
\includegraphics[width=0.49\linewidth]{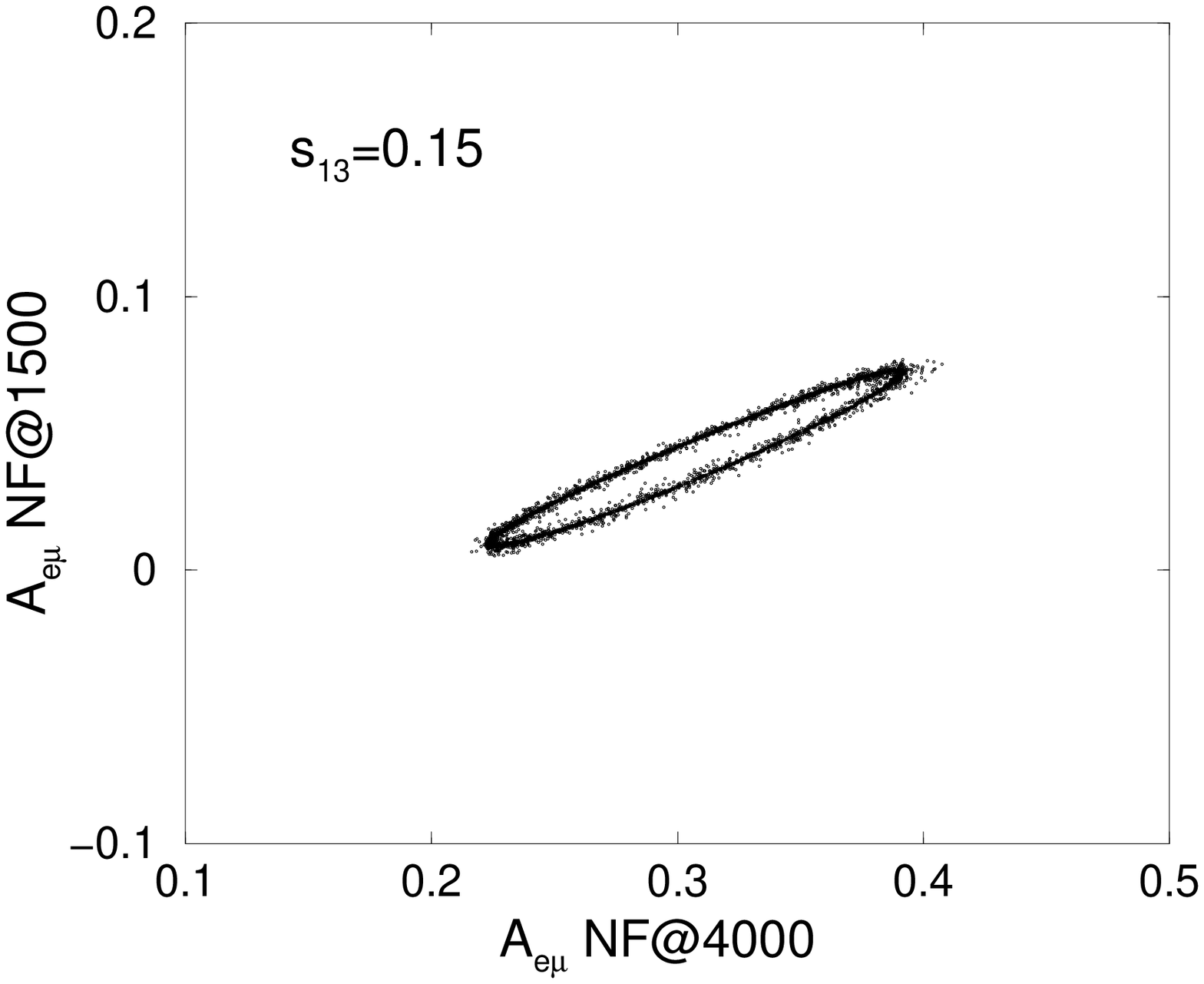}
\caption{\label{fig:AemAem} \it The asymmetry $A_{e\mu}$ computed at two
different baselines $L=4000$ Km (horizontal axis) and $L=1500$ Km
(vertical axis) at the same neutrino energy $E=30$ GeV, for
$\sin{\theta_{13}}= 0.02,~0.05,~0.10,~0.15$. As for the other parameters $\Delta_{31}$ is
 positive,  $\theta_{23}$  is maximal and $\sin^2{\theta_{12}}=1/3$.
}
\end{figure}

For a shorter baseline the sign degeneracy can induce a larger confusion. This
is illustrated in Figs.\ref{fig:deg2}-\ref{fig:degMUV2} that correspond to
Figs.\ref{fig:deg}-\ref{fig:degMUV}
 but with $A_{\mu e}$ computed for $L=295$ Km and $E_\nu=0.75$ GeV  and $A_{e\tau}$
 computed for $L=1500$ Km and $E_\nu=30$ GeV.
In this case, first of all, a better precision would be needed to disentangle the sign ambiguity. But we see that, barring the unfortunate case that the new physics effects are just right to shift the data point from one closed curve to the other one, a meaningful test of the standard picture is still possible.

 \begin{figure}[h!]
\centering 
\includegraphics[width=0.49\linewidth]{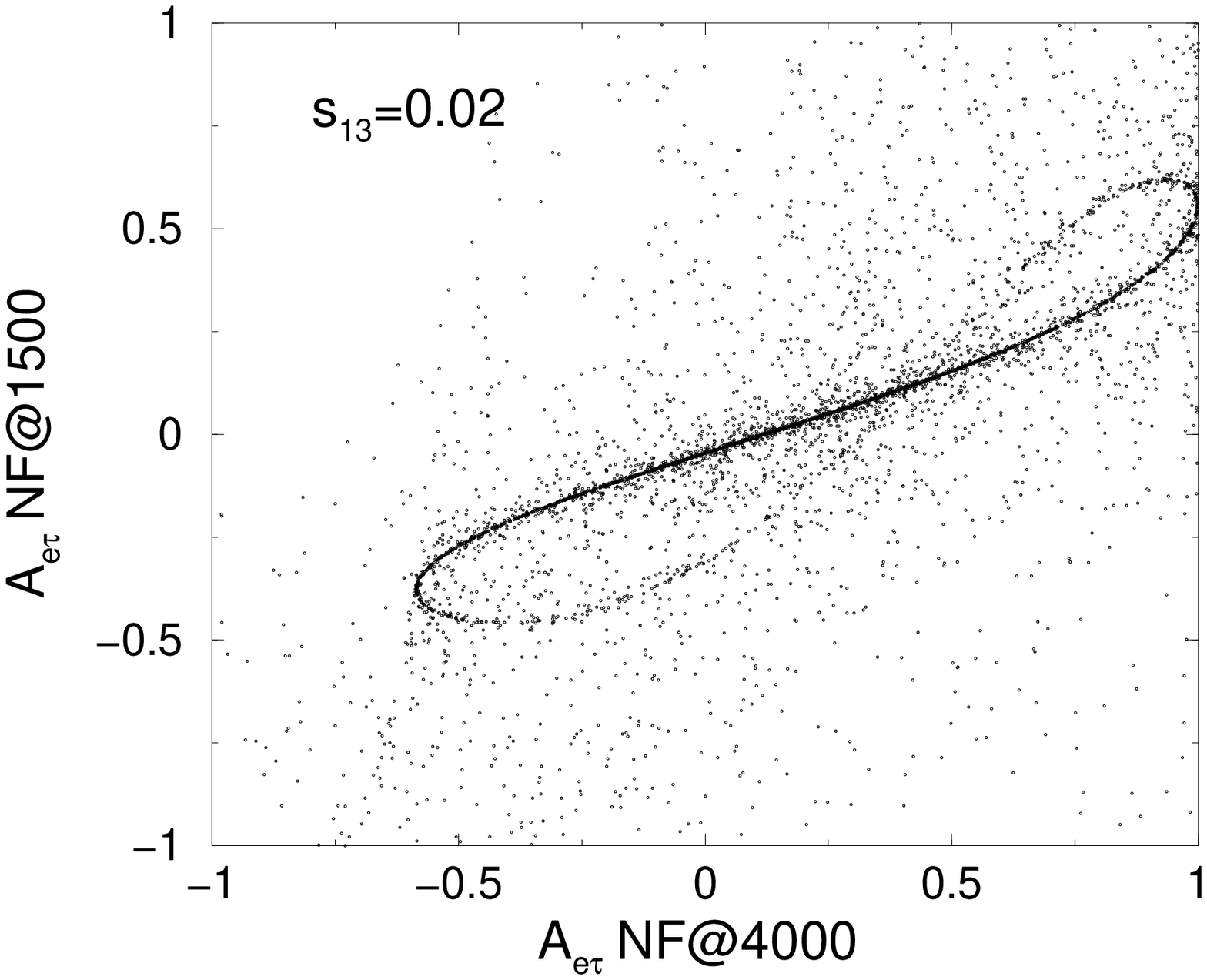}
\includegraphics[width=0.49\linewidth]{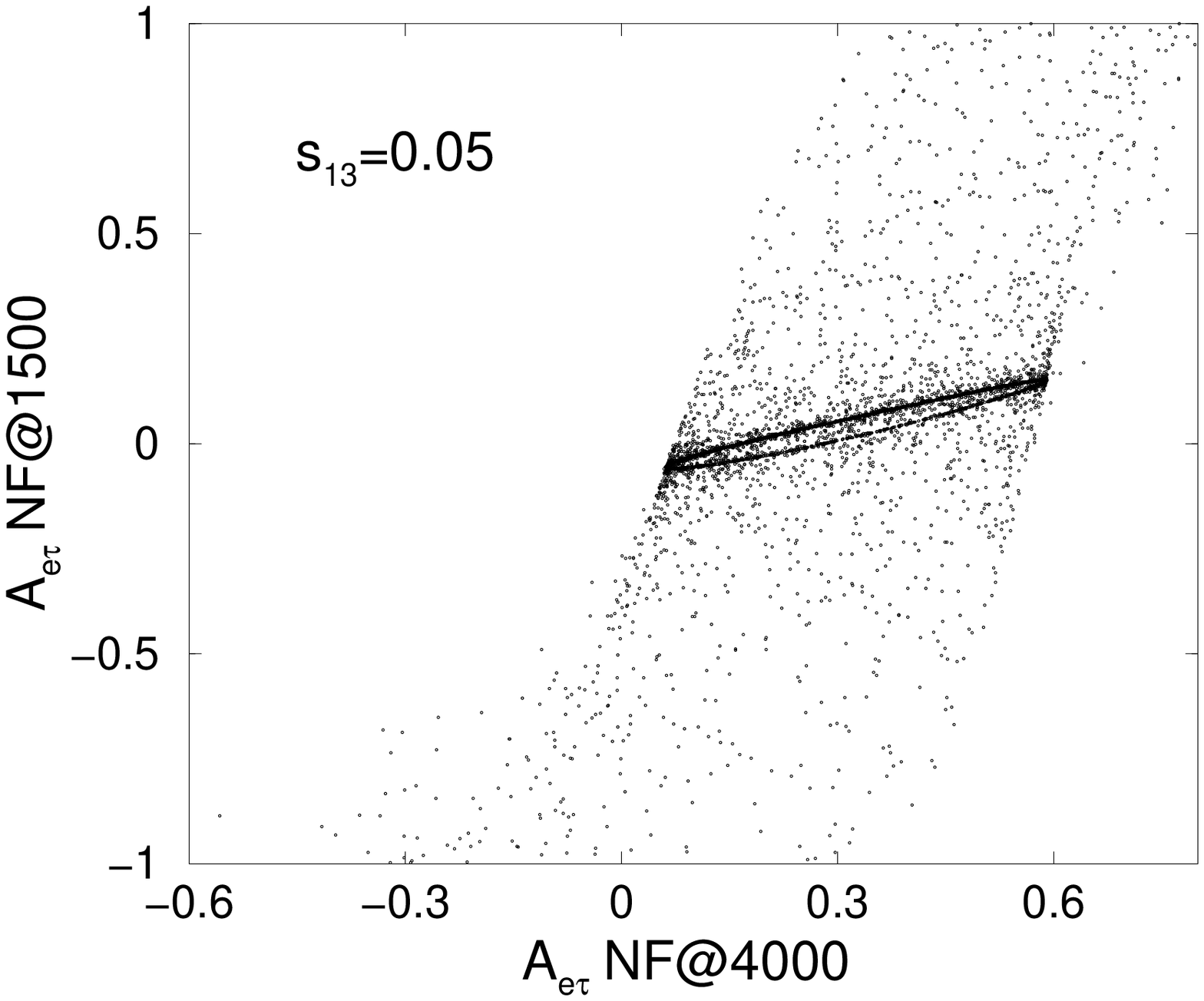}\\
\includegraphics[width=0.49\linewidth]{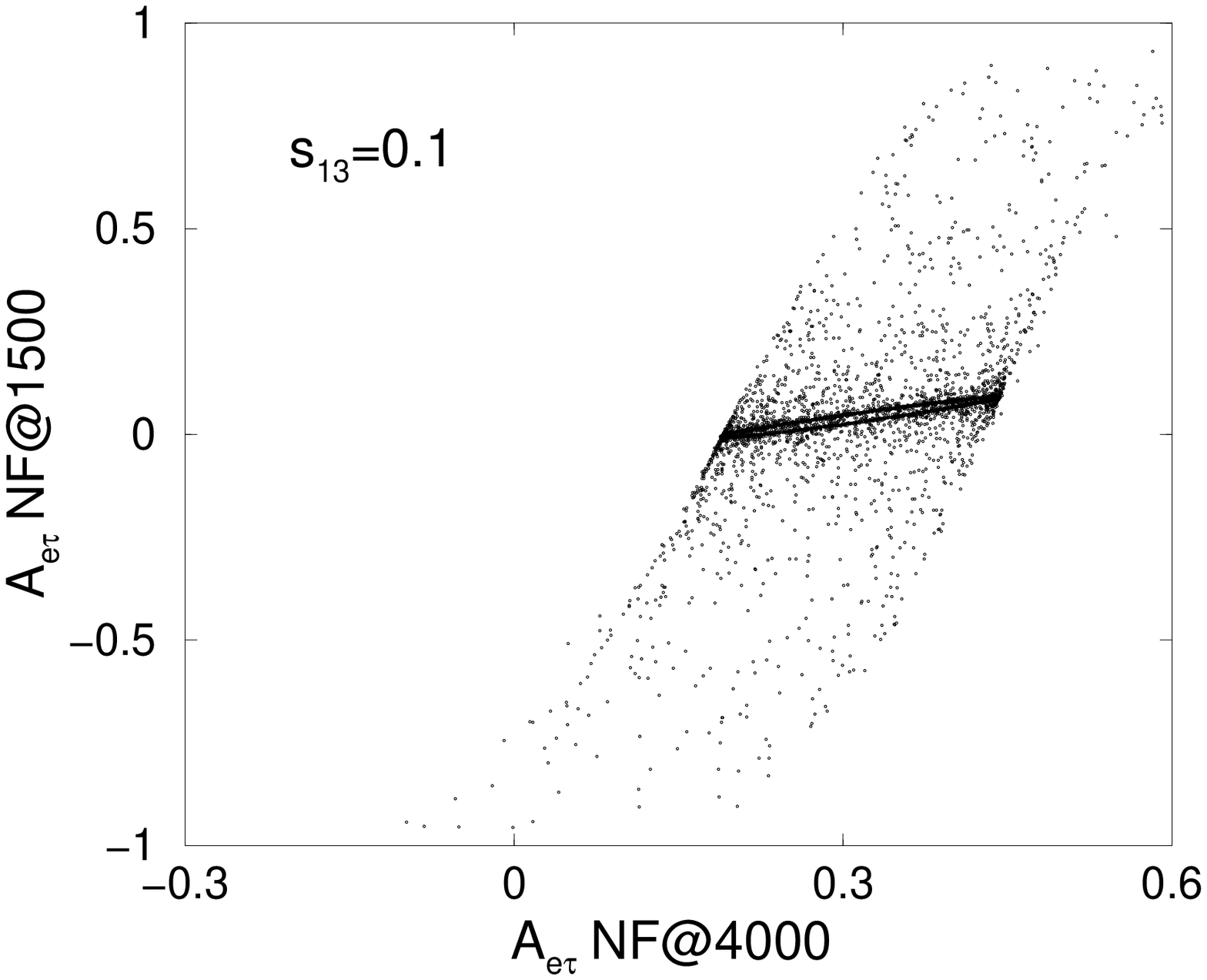}
\includegraphics[width=0.49\linewidth]{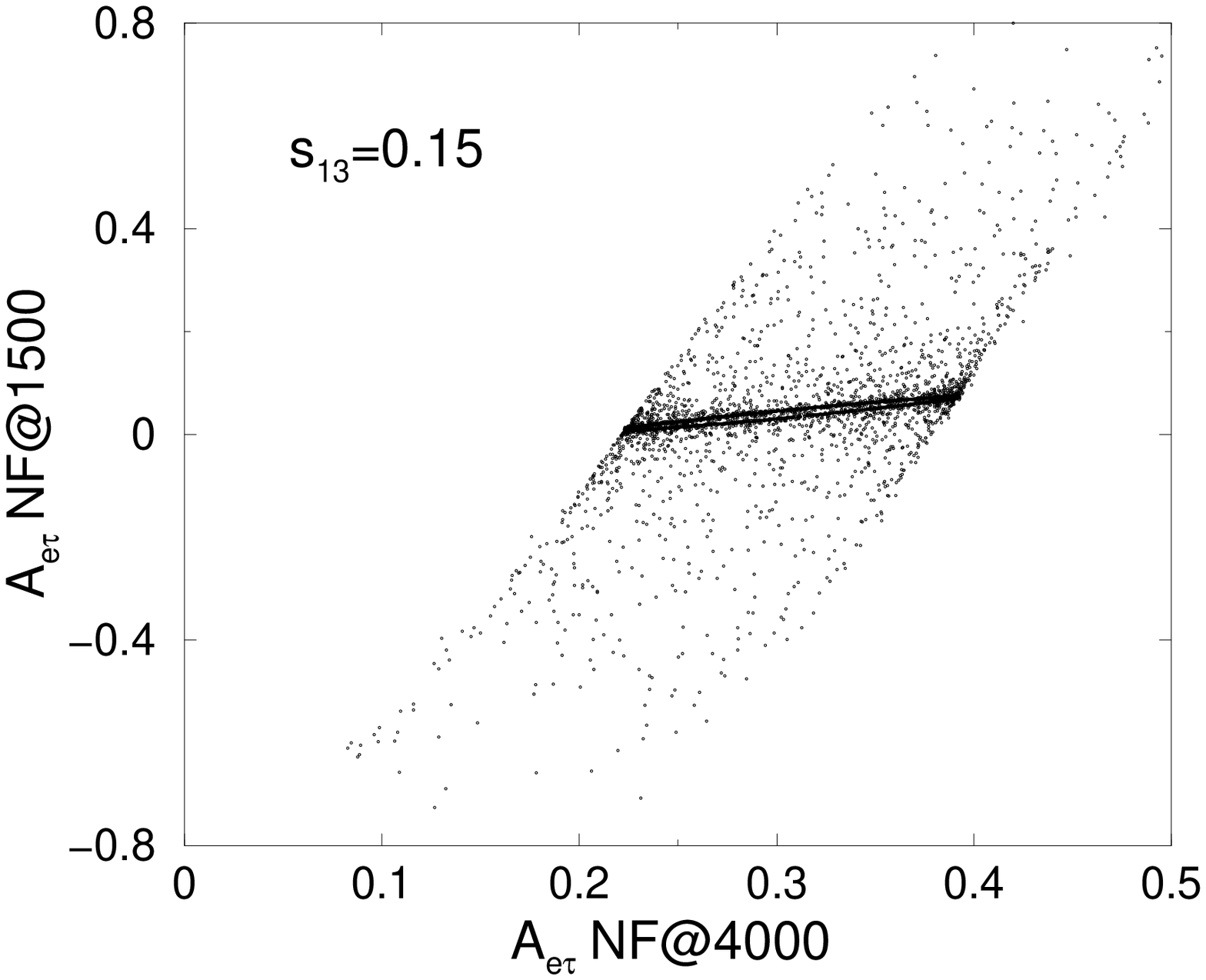}
\caption{\label{fig:AetAet} \it The asymmetry $A_{e\tau}$ computed at two
different baselines $L=4000$ Km (horizontal axis) and $L=1500$ Km
(vertical axis) at the same neutrino energy $E=30$ GeV, for
$\sin{\theta_{13}}= 0.02,~0.05,~0.10,~0.15$.  As for the other parameters $\Delta_{31}$ is
 positive,  $\theta_{23}$  is maximal and $\sin^2{\theta_{12}}=1/3$.
}
\end{figure}

Another complementary possibility of searching for new physics effects that
could also, to some extent, bypass the sign degeneracy
problem is the comparison of a given asymmetry at different baseline distances.
This comparison would be possible in principle even if only the golden channel could be
measured. The plots in Fig.\ref{fig:AemAem}  show $A_{e\mu}$ measured at
$L=4000~{\rm Km}$
and $E_\nu=30~{\rm GeV}$ versus the same asymmetry at $L=1500~{\rm Km}$ and
$E_\nu=30~{\rm GeV}$ for $\sin{\theta_{13}}= 0.02,~0.05,~0.10,~0.15$,  $\Delta_{31}$
 positive,  $\theta_{23}$  maximal and $\sin^2{\theta_{12}}=1/3$.
We see in a
clear way that, for this asymmetry in the MUV case, the possible new physics effects are not sufficiently pronounced
to allow their detection in an experiment with realistic
precision. This confirms that while $A_{e\mu}$ is the best suited asymmetry for a model independent determination of the phase $\delta$, for a test of the standard model one needs to also measure one or more additional asymmetries. The situation is completely different for the $A_{e\tau}$ asymmetry
measured at the two indicated baselines, as shown in Fig.\ref{fig:AetAet}
obtained for the
same values of all parameters. Here the effects of new physics are large enough
to be detectable with reasonable experimental precision.

\section{Conclusion}

In this paper we have considered the proposed facilities for detecting CP violation in neutrino oscillation experiments. All of these experiments aim at measuring the value of $\delta$. We have studied the next question of testing whether the observed CP violation is consistent with the standard model prediction that all CP violation observables in neutrino oscillations must be proportional to the leptonic Jarlskog invariant. As a quantitative model of new physics we have adopted the MUV framework with the present bounds on its parameters.  Although the MUV model is a rather restricted model of new physics (larger effects can be expected in a more general model), we have found that indeed the deviations from the standard picture of CP violation could be detected with realistic experimental accuracies (namely those accuracies which are needed to make a significant measurement of $\delta$ and to disentangle the related sign degeneracy). Actually our results show that the deviations allowed by the present constraints on the MUV parameters are large enough that a meaningful test can be done with a relatively  modest precision.   We have explicitly checked this statement by evaluating the uncertainties associated with the asymmetries $A_{e\mu},A_{e\tau}$  and $A_{\mu\tau}$ as if they were measured at a  neutrino factory with detectors at $L=1500$ Km, showing that new physics effects can produce deviations from the standard model asymmetries well outside of the error bars (in this respect $A_{e\tau}$ is much more promising than $A_{\mu\tau}$ as the asymmetry to be measured in correspondence with $A_{e\mu}$).
But it is also true that the data constraining the new physics parameters will improve in the near future, so that by the time that CP violation can be observed the allowed new physics deviations will presumably be more contained. For testing the PMNS mechanism of CP violation it is necessary to measure not only the golden channel $A_{e\mu}$ (which is the most protected from the new physics effects in the MUV framework) but at least one more channel, for example the silver channel $A_{e\tau}$ (or $A_{\mu\tau}$). One can either measure both channels at the same value of $L/E_\nu$ or the silver channel at two (or more) values of $L/E_\nu$. In the first case, a relatively small value of $L/E_\nu$ is to be preferred, because the ratio of the effects from new vs standard physics in the asymmetries increase with decreasing $L/E_\nu$. Thus $L/E_\nu$ must be sufficiently large to allow a good counting rate for asymmetries, so that CP violation can be observed, and sufficiently small to enhance the visibility of the new physics effects. From our simulations it appears that a good compromise corresponds to a neutrino factory with $L=1500$ Km and $E_\mu \sim 50$ GeV.

%%%%%%%%%%%%%%%%%%%%%%%% ACKNOWLEDGEMENTS   %%%%%%%%%%%%%%%%%%%%%%%%%%%%%%

\section*{Acknowledgements}

We recognize that this work has been partly supported the Italian Ministero dell'Universita' e della Ricerca Scientifica, under the COFIN program for 2007-08.

\appendix
\label{app:intro}
\section{Transition probabilities in the MUV framework}
In this appendix we quote the vacuum probabilities obtained in the MUV framework (and in the standard model if not previously quoted in the main text) using the approximations described in Sects.\ref{SM} and   \ref{summuv}.
\subsection{$P_{e\mu}$}
\label{app:emu}
The transition probability  $P_{e\mu}$ reads:
\bea
\nn
P_{e\mu} &=& P_{e\mu}^{\rm SM}+\eta_{e\mu}\left\{r\,\sin\Delta_{31}\left[3\sin(\delta-\Delta_{31}-\delta_{e\mu})+
\sin(\delta+\Delta_{31}-\delta_{e\mu})\right]\right. \\ &&
+\frac{2}{3}\Delta_{21}
\left[\sin(2\Delta_{31}-\delta_{e\mu})-3
\sin \delta_{e\mu}\right]\}
+ 2\eta_{e\tau}  \left\{r\,\cos(\delta-\delta_{e\tau})\sin^2\Delta_{31}
\right.        \\ && \nn
+\frac{2}{3}  \Delta_{21} \cos(\Delta_{31}-\delta_{e\tau}) \sin\Delta_{31}
\left. \right\}  + \\ \nn && \eta_{e\mu}\eta_{e\tau}\sin \Delta_{31}
\left[-3\sin(\Delta_{31}-\delta_{e\tau}+\delta_{e\mu})+
\sin(\Delta_{31}-\delta_{e\mu}+\delta_{e\tau})\right]  +       \\ \nn &&
\frac{1}{2}\eta_{e\mu}^2(5+3\cos 2\Delta_{31})+
 \eta_{e\tau}^2   \sin^2 \Delta_{31}.
\eea 

Let us analyze the new CP violating terms, namely those linearly suppressed by the new parameters $\eta_{e\mu}$ and $\eta_{e\tau}$ (we can safely assume that the second-order term   $\eta_{e\mu}\,\eta_{e\tau}$ is smaller than the others). In each curly brackets there is a competition between the term suppressed by $\theta_{13}$ and that suppressed by $\Delta_{21}$; since $\Delta_{21}\sim \mathcal O(10^{-4})\, L/E_\nu$, for small enough $L/E_\nu$ it can be neglected, unless $r$ is of the same order of magnitude (in fact, this is almost the regime in which we work because  $L /E_\nu$ is at most $\mathcal O(10^{3})$ Km/GeV).      For the coefficient of $\eta_{e\tau}$          this means that we are left with a CP conserving term and it does not contribute to CP violation in the $e\mu$ channel. On the other hand, for
  $\eta_{e\mu}$, the relevant term is CP violating and only linearly suppressed with $\theta_{13}$. We recover here the observation made in \cite{gavela} that at short enough baselines, the new CP violating terms in $P_{\alpha\beta}$ mainly depend on the phase associated to the corresponding $\eta_{\alpha\beta}$.    This also means that, due to the strong bounds on $\eta_{e\mu}$ in MUV, $A_{e\mu}$ is a good CP violating observable to measure the standard phase $\delta$.

\subsection{$P_{e\tau}$}
\label{app:etau}
The  MUV effects in the $P_{e\tau}$  probability are obtained from $P_{e\mu}$ of eq.(\ref{app:emu}) changing $\eta_{e\mu} \leftrightarrow \eta_{e\tau}$    and $\delta_{e\mu}\leftrightarrow \delta_{e\tau}$:
\bea
\nn
P_{e\tau} &=& P_{e\tau}^{\rm SM}+\eta_{e\tau}\left\{r\,\sin\Delta_{31}\left[3\sin(\delta-\Delta_{31}-\delta_{e\tau})+
\sin(\delta+\Delta_{31}-\delta_{e\tau})\right]\right. \\ &&
+\frac{2}{3}\Delta_{21}
\left[\sin(2\Delta_{31}-\delta_{e\tau})-3
\sin \delta_{e\tau}\right]\}
+ 2\eta_{e\mu}  \left\{r\,\cos(\delta-\delta_{e\mu})\sin^2\Delta_{31}
\right.        \\ && \nn
+\frac{2}{3}  \Delta_{21} \cos(\Delta_{31}-\delta_{e\mu}) \sin\Delta_{31}
\left. \right\}  + \\ \nn && \eta_{e\mu}\eta_{e\tau}\sin \Delta_{31}
\left[-3\sin(\Delta_{31}+\delta_{e\tau}-\delta_{e\mu})+
\sin(\Delta_{31}+\delta_{e\mu}-\delta_{e\tau})\right]  +       \\ \nn &&
\frac{1}{2}\eta_{e\tau}^2(5+3\cos 2\Delta_{31})+
 \eta_{e\mu}^2   \sin^2 \Delta_{31}.
\eea 

We can apply to  $P_{e\tau}$ the same considerations as for $P_{e\mu}$, concluding that the relevant new CP violating effects are all contained in the terms proportional to $\eta_{e\tau}$.

\subsection{$P_{\mu\tau}$}
\label{app:mutau}

The probability reads: 
\bea
\nn
P_{\mu \tau} &=& P_{\mu\tau}^{\rm SM}-2   \eta_{\mu\tau}    \sin \delta_{\mu\tau}
\left(\sin 2\Delta_{31}-\frac{4}{3}\Delta_{21}\cos2\Delta_{31}\right) + \\ 
&& 2\eta_{e\tau}\left[r \cos(\delta-\delta_{e\tau})\sin^2\Delta_{31}+\frac{2}{3}\Delta_{21}\cos(\Delta_{31}+\delta_{e\tau})
\sin \Delta_{31}\right]  + \\ \nn && \eta_{\mu\tau}
\sin \delta_{\mu\tau} \sin 2\Delta_{31} (\eta_{\mu\mu}+\eta_{\tau\tau})
\eea

where the standard model expression is given by:

\bea
P_{\mu\tau}^{\rm SM} &=& \sin^2\Delta_{31} + \frac{4}{3}\Delta_{21}\sin\Delta_{31}(r \sin \delta \sin \Delta_{31}+s
\cos\Delta_{31})   + \sin^2\Delta_{31} (4a^2-r^2) \\
 \nn && -\frac{2}{3}\Delta_{21}\sin2\Delta_{31}-
\frac{2}{9}\Delta_{21}^2 (1-3\cos 2\Delta_{31}).
\label{aa}
\eea

Contrary to    $P_{e\mu}$       and    $P_{e\tau}$, the larger new physics effect in   $P_{\mu \tau}$ is only linearly suppressed by $\eta_{\mu\tau}$. This allows to approximate the probability as in eq.(\ref{eq:pmutau}) and to get the simple expression for the asymmetry $A_{\mu\tau}$ as given in eq.(\ref{eq:Amutau}).

\subsection{$P_{\mu\mu}$}
\label{app:mumu}
The expression of the transition probability  $P_{\mu\mu}$ reads:
\bea
P_{\mu \mu} &=& P_{\mu \mu}^{\rm SM}+ 4\,\eta_{\mu\tau} \cos\delta_{\mu\tau}\,\sin^2 \Delta_{31}\,(2\,a+\eta_{\mu\tau}\,\cos \delta_{\mu\tau})
\label{ese}
\eea
where

\bea
\nn
P_{\mu \mu}^{\rm SM}&=& 1-(1-4\,a^2)\,\sin^2 \Delta_{31}-\frac{2}{9}\,\Delta^2_{21}\,(1+3\cos 2\Delta_{31})
+\frac{2}{3}\,\Delta_{21}\,\sin 2\Delta_{31}\,(1-s-r\cos \delta).\\&&
\label{essse}
\eea                                                                   
 
\vfill

%%%%%%%%%%%%%%%%%%%%%%%% REFERENCES   %%%%%%%%%%%%%%%%%%%%%%%%%%%%%%

\end{document}